\documentclass[prb,twocolumn,showpacs]{revtex4-1}
\pdfoutput=1
\usepackage{amssymb,latexsym}
\usepackage{amsmath,amsbsy,bbm}
\usepackage{ifpdf}
\usepackage{epsfig,bm}
\usepackage{graphicx,comment}
\usepackage{color}
\usepackage{soul}
\usepackage[normalem]{ulem}
\begin{document}

\title{Applications of neural networks to the studies of phase transitions of 
two-dimensional Potts models}
\author{C.-D. Li}
\affiliation{Department of Physics, National Taiwan Normal University,
88, Sec.4, Ting-Chou Rd., Taipei 116, Taiwan}
\author{D.-R. Tan}
\affiliation{Department of Physics, National Taiwan Normal University,
88, Sec.4, Ting-Chou Rd., Taipei 116, Taiwan}
\author{F.-J. Jiang}
\email[]{fjjiang@ntnu.edu.tw}
\affiliation{Department of Physics, National Taiwan Normal University,
88, Sec.4, Ting-Chou Rd., Taipei 116, Taiwan}
%\vspace{-2cm}

\begin{abstract}
We study the phase transitions of two-dimensional (2D)
$Q$-states Potts models on the square lattice, using the first principles 
Monte Carlo (MC) 
simulations as well as the techniques of neural networks (NN). 
We demonstrate that the ideas from NN can be adopted to study these 
considered phase transitions efficiently. In particular, even with a 
simple NN constructed in this investigation, we are able to obtain the 
relevant information of the nature of
these phase transitions, namely whether they are first order or second 
order. Our results strengthen the potential 
applicability of machine learning in studying various states of matters. 
Subtlety of applying NN techniques to investigate many-body systems
is briefly discussed as well.
  
\end{abstract}

%\pacs{12.39.Fe, 75.10.Jm, 75.40.Mg, 75.50.Ee}

\maketitle

\section{Introduction}
%{\bf Introduction} --- 
While originally being a topic in computer science such as
the artificial intelligence,
the techniques of machine learning (ML) have been applied to many other
fields of science in the last a few decades \cite{Has09,Hor11,Gyo12,Kir14,Kel15,Lib15,Goo16,Wil16}. 
For example, 
the strategy of using methods of ML in analyzing large data sets of traffic 
flow and genome sequences has gained a lot of attention recently.
Ideas from ML provide
alternatives to the conventional approaches in various scientific
areas. In some cases, the performance of these new approaches
using the techniques of ML is as efficient as, or even better than the 
conventional ones. In the near future, it is anticipated that ML is 
likely to play a crucial role in many traditional fields of science.

The theoretical study of different phases of many-body systems, as well as 
the characters of the transitions between these various states is one of the 
major research
topics in condensed matter physics \cite{Sac99}. 
Numerous numerical techniques, such
as Monte Carlo simulations, series expansion, exact diagonalization, 
tensor networks and so on, are available for fulfilling these tasks 
\cite{San10,Ave13}. Indeed, studies using these traditional methods have provided 
valuable information regarding the properties of matters. In particular,
using the theoretical results obtained by these conventional techniques 
in conjunction with the data of relevant experiments, properties of some materials 
have been understood to an unprecedented precision.

With the advance of computing power as well as the development of 
ML algorithms, one interesting question arises, namely are these techniques of
ML capable of efficiently studying different phases of matters? Such an idea
has triggered pioneering works of applying ML algorithms to investigate 
phases, particularly the transitions between different states, of 
certain condensed matter systems
\cite{Car16,Tro16,Tor16,Bro16,Chn16,Tan16,Nie16,Liu16,Xu16,Wan16,Wan17,Liu17,Che17}. 
In these pioneering studies, it is 
demonstrated that the neural networks (NN) of ML is a powerful
approach to distinguish phases of many-body systems. 
Furthermore, NN can even detect exotic states such as a phase with 
topological order. Finally, NN is capable of locating the place in
the relevant parameter space where the transition between two phases 
occurs as well.

Motivated by the success of applying NN to study phases of condensed matter 
systems, in this investigation we examine if one can obtain further 
information regarding a phase transition beyond detecting the location of 
the critical point. For instance, it will be extremely interesting to determine 
whether a studied phase transition is continuous or not,
by employing either the supervised or unsupervised NN 
techniques \cite{unsupervised}.
Notice for the studies of using supervised NN to 
investigate the phase transitions of condensed matter systems, the 
considered NN typically consists of one input layer, some hidden 
layers, including the convolutional and pooling layers, and one output layer as 
well as the weights connecting different layers. The constructed NN is then 
trained by feeding it with  
configurations of the two phases associated with the studied phase transition.
Furthermore, in the 
training stage, a particular value (number) is assigned to each configuration 
as its default (and fixed) result in the output layer, and the weights are 
calculated by minimizing 
the so called loss functions during the training process. 
When the training is done, one can use
the trained NN to obtained the outputs of the target
configurations (called the testing set). These outputs are functions of the 
considered parameters and they provide us with relevant information regarding 
the critical points of the investigated phase transitions. For instance,
it is shown convincingly that for the Ising model (on the square lattice), 
as the box sizes $L$
increase, the corresponding temperatures at which 
both the components of the output vectors are 0.5 converge 
to the correct critical temperature \cite{Car16}.

While the outputs described in previous paragraph may be useful in determining 
whether the studied phase transitions are first order or second order, in this
investigation we consider the possibility of using another powerful and 
completely different idea, implemented in
the NN framework, to decide the nature of a phase transition. Specifically, 
in our study we use the histogram method which is one of  
the most frequently considered methods in the
conventional approaches. The histogram method will be introduced briefly
later in the related section. Remarkably, for the considered 
models here, namely the two-dimensional (2D) 
$Q$-states Potts models with $Q$ being a positive integer 
\cite{Bin81,Wu82,Bil95}, even a simple NN in conjunction with the histogram 
method can provide us with the relevant information to unambiguously determine 
the nature of the phase transitions, induced by tuning the temperatures, 
of these models. In particular, unlike the typical supervised NN approaches,
in this study we confuse the built NN here by using only the configurations of the 
ordered phases as the training sets. As we will demonstrate later, this idea leads 
to efficient determination of the critical points in the relevant parameter
spaces as well as the nature of the considered phase transitions.
Our investigation is not only important and interesting in 
itself, the obtained results also strengthen the applicability of NN in 
studying condensed matter systems.

\section{Microscopic models, observables, and methods}
%{\bf Microscopic models, Observables, and Methods} --- 
The Hamiltonian $H$ of 
the 2D $Q$-states Potts model on the square lattice considered in our study is 
given by
\begin{equation}
\beta H = -\beta \sum_{\left< ij\right>} \delta_{\sigma_i,\sigma_j},
\end{equation} 
where $\beta$ is the inverse temperature and $\left< ij \right>$ stands for the 
nearest neighboring sites $i$ and 
$j$. In addition, the Potts variable $\sigma_i$ appearing above at each site 
$i$ has the expression
\begin{equation}
\sigma_j = \exp\left(i\frac{2\pi s_j}{Q}\right), s_j = 1,2,3,...,Q.
\end{equation}
In the literature, for the 2D $Q$-states Potts models on the square lattice,
the phase transitions due to fluctuation in temperatures are second order for $Q\le 4$, and
first order when $Q \ge 5$.
In this investigation, we have studied these
phase transitions using both the method of Monte Carlo (MC) calculations 
and the techniques of NN. In particular, special attention 
has been focused on whether the methods of NN can be employed to determine the 
nature of these considered phase transitions efficiently.     

To study these phase transitions through the MC simulations, 
the observable magnetization per site 
$\left< |m|\right>$, where $m$ takes the following form
\begin{equation}
m = \frac{1}{L^2} \sum_{i}\sigma_i,
\end{equation}   
is measured. Here $L$ is the linear box sizes used in the simulations. 
Besides, the energy density (energy per site) $\left< E\right>$ is calculated 
as well. By studying the histograms of these two physical quantities, the 
nature of these phase transitions, i.e., first order or second order, can be 
effectively decided.

In addition to the conventional approach of using MC simulations,
we also investigate these phase transitions with the techniques of
supervised NN. By considering quantities which are defined within the NN 
framework, we find that the nature of these phase transitions can be accurately
determined as well. The procedures, including the considered quantities, 
associated with the constructed NN in this study will be described in details 
in the relevant sections.

\section{The numerical results}
%{\bf The Numerical Results} --- 
To investigate the nature of the considered phase transitions, we
have performed MC simulations for $Q=2,3,4,5$, and 10, using
the Swendsen Wang algorithm \cite{Swe87,New99,Lan14}. In particular, at every
considered temperature close to the critical temperatures $T_c(Q,L)$, a few 
thousand configurations are stored for each studied $Q$ and $L$. 
These configurations will be used in the 
NN approach to construct the appropriate quantities serving as the
main information of deciding the nature of the considered phase transitions. 
In the next subsection, we will briefly present the numerical results obtained
from the MC simulations.

\subsection{The results of MC Simulations}
%{\it Results of MC Simulations} --- 
One efficient method to determine the nature of a phase transition,
namely to decide whether the transition is first order or second order,
is to investigate the histograms of some relevant observables. Specifically,
if a phase transition is first order, then for sufficiently large  
$L$, two peaks will appear in the histograms of certain observables
when the simulations are carried out at (or close to) the corresponding 
critical point. Furthermore, the relative heights between the
peak(s) and the trough (located in the middle) will approach infinity as one
increases $L$. The appearance of the two peaks scenario for 
a first order phase transition is due to the fact that at the critical point
the system is in either phase with equal probabilities. 
On the other hand, if the considered phase transition is 
second order, then the related histograms do not have these features.
It should be pointed out that while for a second order phase transition,
one may observe the appearance of two peaks in the histograms (of
certain observables), the mentioned relative heights between
the peak(s) and the trough will saturate to a constant.
With this powerful idea, we indeed find that while the phase transitions
for 2D Potts models on the square lattice with $Q=2,3$, and 4 are continuous, 
the corresponding transitions of $Q=5$ and 10 are first order. 

\begin{figure}
%\vskip0.5cm
\begin{center}
\vbox{
\includegraphics[width=0.325\textwidth]{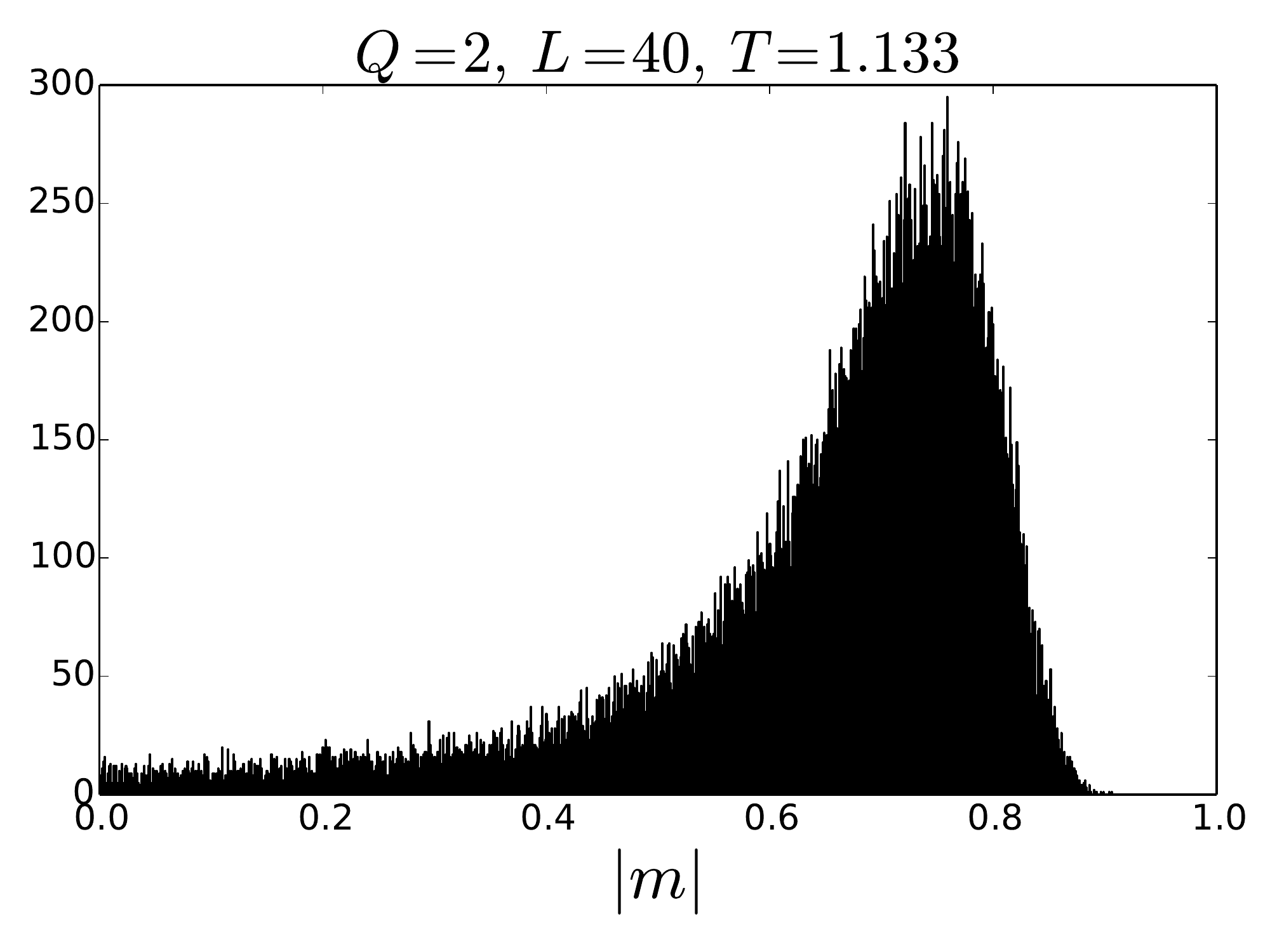}\vskip0.2cm
\includegraphics[width=0.325\textwidth]{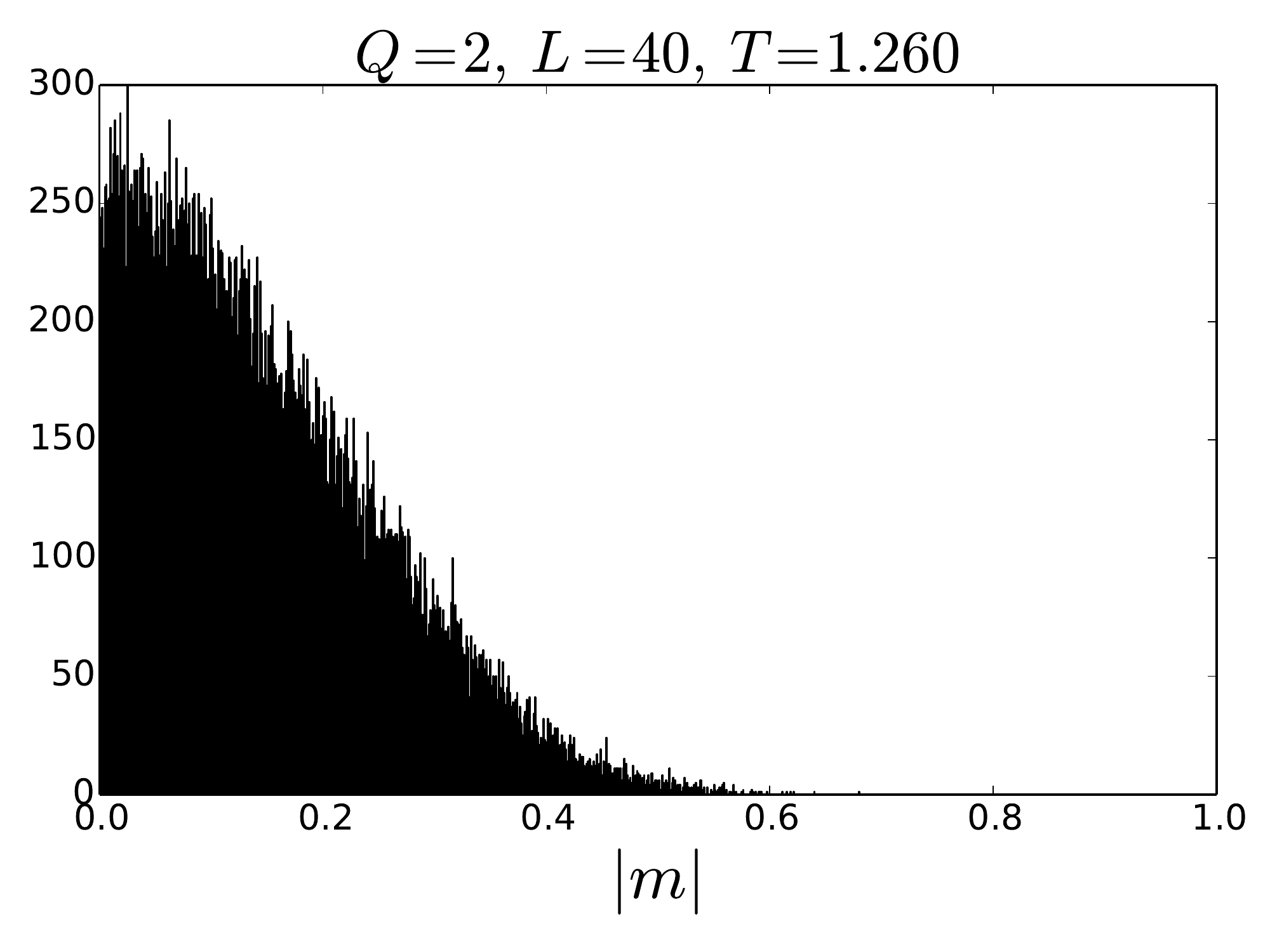}
}
\end{center}\vskip-0.7cm
\caption{Histograms of $\left< |m|\right>$ at $T=1.133$ (top)
and $T=1.260$ (bottom) for 2D $2$-states Potts model with $L = 40$ 
on the square lattice. Several ten thousand data, separating by few thousand 
updates from each other after the thermalization, are generated for each of 
the plots. The histograms are produced by the ``hist'' function of 
pylab \cite{pyl}.}
\label{MC_1}
\end{figure}

The histograms of $\left< | m |\right>$ for $2$-states Potts models determined with $L = 40$
are shown in fig.~\ref{MC_1} \cite{histograms}.  
In particular, the results in fig.~\ref{MC_1} are obtained at temperatures 
1.133 
(top) and 1.260 (bottom). We find that no two
peaks structure shows up for other temperatures, separating from each other 
by 0.001, between $T = 1.133$ and $T = 1.260$. We have 
additionally simulated $L = 80$ lattices for 2-states Potts model and have 
found a similar scenario as that of $L = 40$. Based on these outcomes,
one concludes that the phase transition of 2-states Potts model is second
order. 

\begin{figure}
%\vskip0.5cm
\begin{center}
\vbox{
\includegraphics[width=0.325\textwidth]{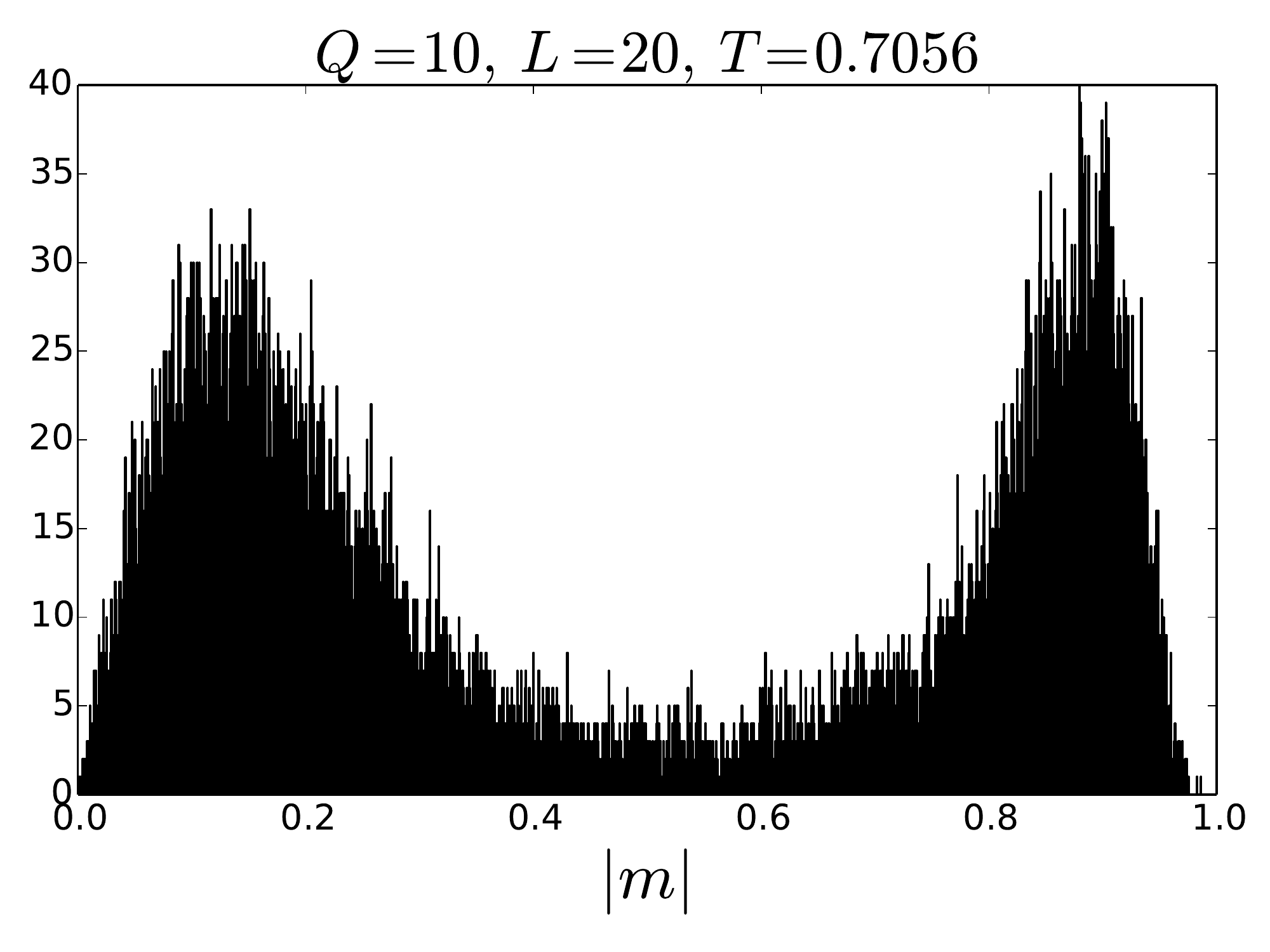}\vskip0.2cm
\includegraphics[width=0.325\textwidth]{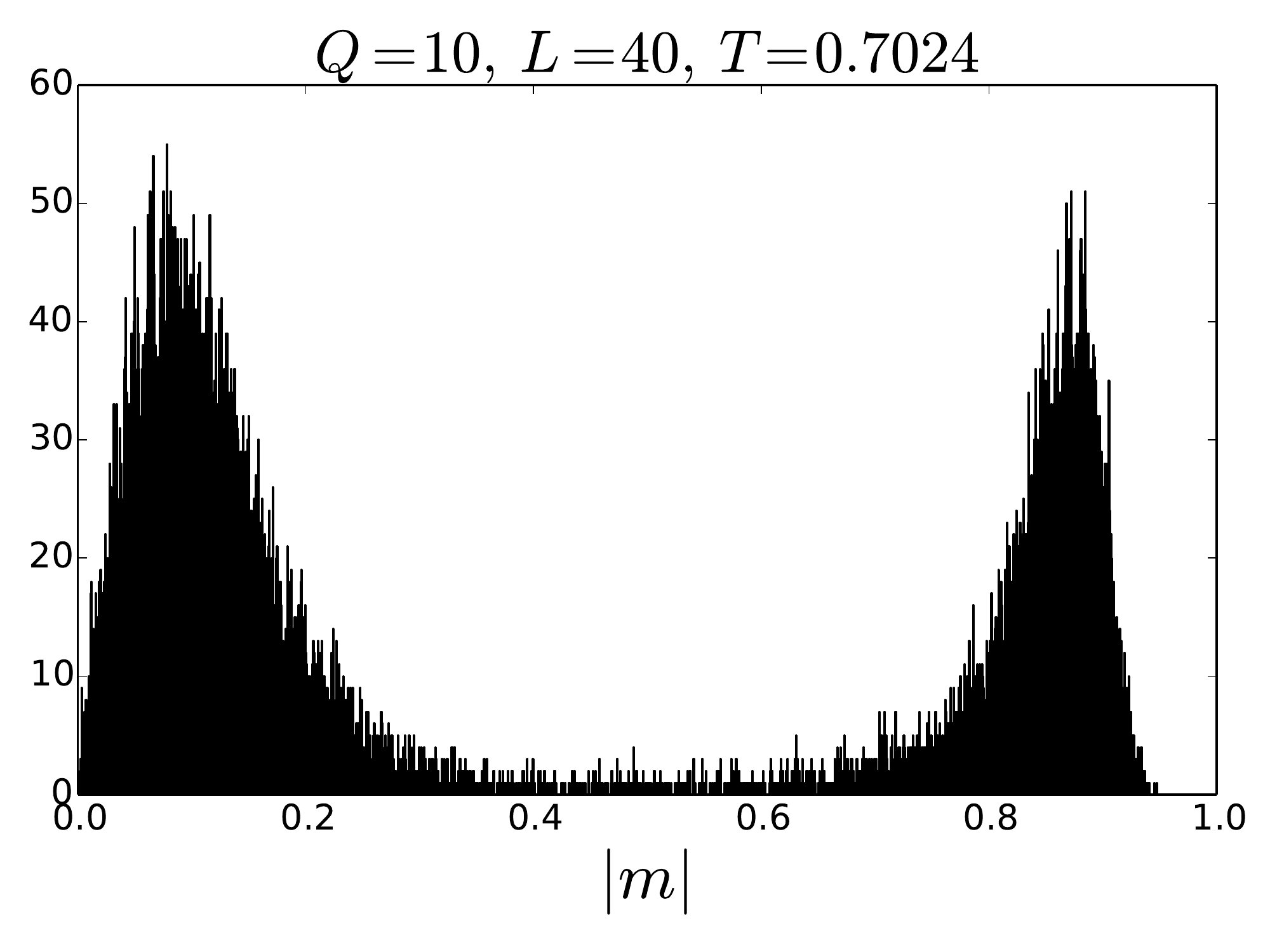}
}
\end{center}\vskip-0.7cm
\caption{Histograms of $\left< |m|\right>$ for $L = 20$ (top)
and 40 (bottom) for 2D $10$-states Potts model on the square lattice. 
Around twenty thousand data, separating by few thousand updates from each 
other after 
the thermalization, are generated for each of the plots. The histograms are 
produced by the ``hist'' function of pylab \cite{pyl}.}
\label{MC_2}
\end{figure}

The histograms of $\left< | m |\right>$ for $10$-states Potts models,
determined at $L=20$ and $L=40$, are demonstrated  
in fig.~\ref{MC_2}. Notice two peaks clearly appear for both $L=20$
and $L = 40$ in fig.~\ref{MC_2}.
In particular, this two peaks structure is more noticeable for $L=40$.
These results indicate that the phase transition for $10$-states Potts model 
is indeed first order. 

To summarize, figs.~\ref{MC_1} and \ref{MC_2} clearly 
show convincing signals that the phase transitions of $2$- and $10$-states 
Potts models are second order and first order, respectively.  
It should be pointed out that because 
$Q=4$ and $Q=5$ are at the edges of separating second and first order phase 
transitions of 2D Potts models on the square lattice, data of larger $L$ are 
needed in order to reach a conclusive answer of whether the phase 
transitions of $4$- and $5$-states Potts models are second order or first 
order. Although our data for $4$- and $5$-states Potts model do not lead to 
absolutely conclusive results, they do show the correct trends. 

\subsection{The Results of NN Methods}
%{\it Results of NN Methods} --- 
Besides the conventional methods, as the one of Monte Carlo simulations shown 
in the previous subsection, in the following we will employ the techniques of supervised ML 
to study the considered phase transitions.
  
\subsubsection{The training sets, labels, and outputs of NN}
Before presenting the numerical results obtained using the techniques of
supervised ML, 
we will firstly detail the ideas behind our NN approach. In particular,
the considered NN training sets and labels will be introduced thoroughly. 
To begin with, a typical NN procedure of studying the phase
transition of Ising model will be described briefly.

When one investigates the phase
transition of Ising model with the supervised NN techniques, the training set
consists of configurations obtained by Monte Carlo simulations
performed over a broad range of temperatures. Furthermore, these selected 
temperatures 
should be from both sides of the critical temperature $T_c$.
The label of a configuration in the training set is an assigned
two-component vector, taking the result of either $\left(1,0\right)$ or $\left(0,1\right)$.
These vectors, namely $\left(1,0\right)$ and $\left(0,1\right)$, serve as the flags to indicate 
whether the configurations are in the disordered phase or the ordered phase. Notice the training 
is conducted for every considered lattice with linear size $L$. After the completion of  
training, one then uses the trained NN to calculate the outputs of many testing configurations 
obtained with temperatures crossing $T_c$. In this testing stage, for each linear lattice 
size $L$, the corresponding critical temperature $T_c(L)$ is the temperature at which the output 
vector is $\left(0.5,0.5\right)$.

Having briefly introduced a typical procedure of studying the phase transition of
Ising model using the NN methods, in the following we will specify our NN approach
to the investigation of $Q$-states Potts models. In particular, the ideas behind the constructed
NN used in our study will be explained in detail. Notice
for $Q$-states Potts model on a $L$ by $L$ lattice, one intuitively expects that  
the corresponding ground states consist of $Q$ configurations. In addition, 
for the $j$th ground state configuration, the associated Potts variable at 
every site has the same positive integer value $j$. On the other hand,
at temperatures $T$ much higher than the critical temperature $T_c$,
the corresponding Potts variable at each site of a $L$ by $L$ lattice can take any number 
$n_2$ in $\{1,2,3,...,Q\}$ with equal probability. These two propositions
are confirmed by the snapshots of our Monte Carlo simulations. Based on these results,
instead of performing our study following the related standard NN procedures,
here we consider a completely different approach which is much simpler conceptionally. 
Specifically, in our NN calculations, for any $Q$-states Potts model we use 
the associated $Q$ (ground state) configurations as the objects in the 
pre-training set (We will explain later why we call these configurations the 
pre-training set). In other words, unlike the typical 
NN approaches, the training sets considered here are not generated numerically by Monte 
Carlo simulations. On the contrary,
they are created by hand and are the expected ground state configurations of the studied 
models.

\begin{figure}
%\vskip0.5cm
\begin{center}
\includegraphics[width=0.4\textwidth]{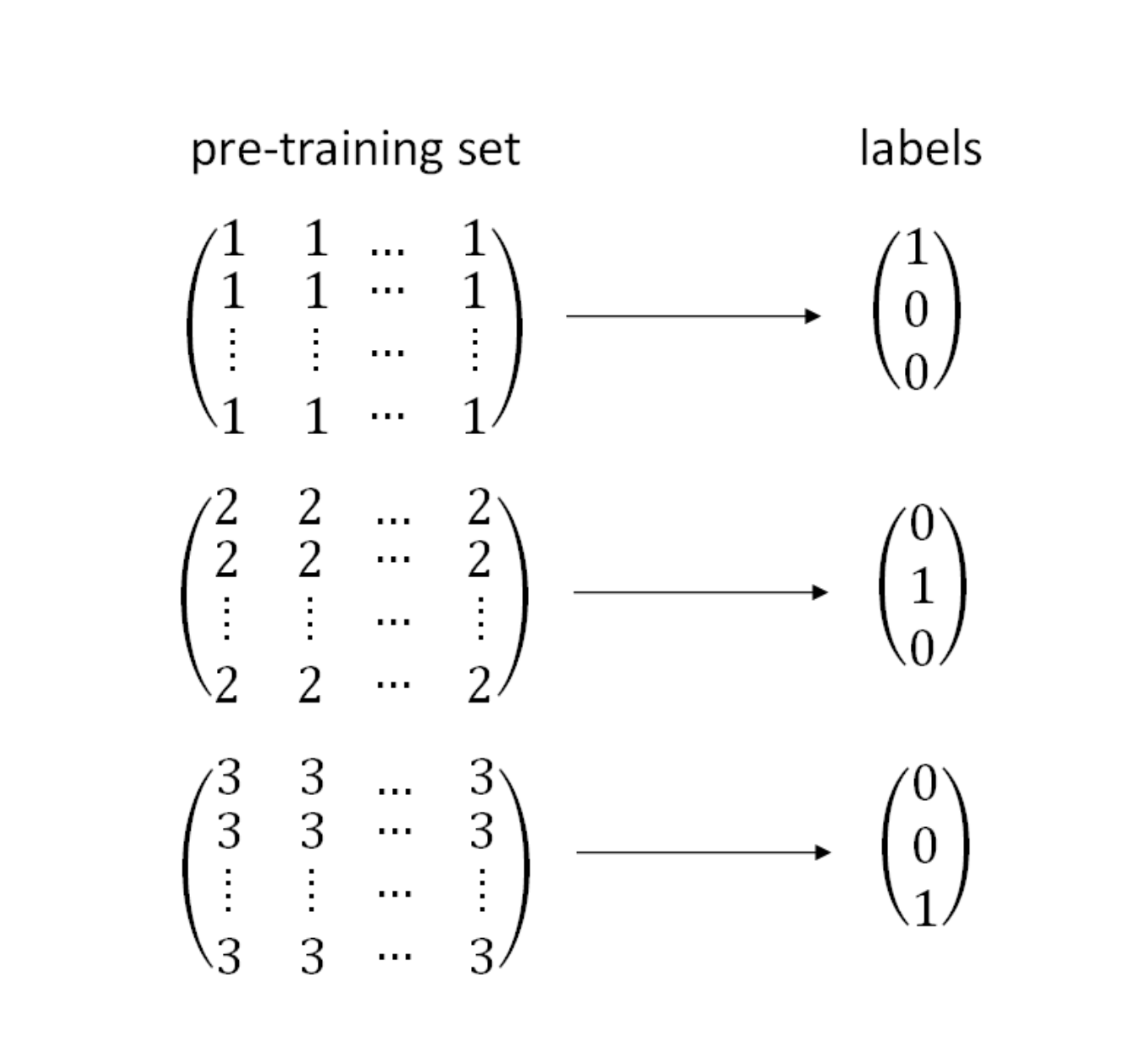}
\end{center}\vskip-0.7cm
%\vskip2.0cm
\caption{The pre-training set and their corresponding NN labels for 
3-states Potts model. Each possible ground state configuration on a $L$ by $L$
lattice is represented by a $L$ by $L$ matrix. In particular, the matrix 
entries are the values of the Potts variables at the corresponding lattice
sites. Such a correspondence between
the (ground state) configurations in the pre-training set and their labels 
can be applied to other $Q$-states Potts models in exactly the same manner.}
\label{NN_0}
\end{figure}

Notice for the Ising model, if one uses ordered and disordered states as the two categories 
(classifications) for the training set, 
then the corresponding NN labels are two-component vectors. 
In particular, during the training stage the labels for the ordered and disordered states
are $(1,0)$ and $(0,1)$ (or vice versa), respectively. Inspired by this, we consider $Q$-vectors 
as the NN labels (and outputs) for $Q$-states Potts model. Specifically, in the training process,
a (ground state) configuration on a $L$ by $L$ lattice with each Potts variable taking the same integer value $n_1$ 
is assigned a $Q$-vector as its label. In addition, this $Q$-vector has the 
property that all its components 
are $0$ except the $n_1$th component which is $1$. Figure \ref{NN_0} shows the connection
between the objects in the pre-training set and their corresponding labels, namely $Q$-vectors 
for 3-states Potts model. 
The $L$ by $L$ matrices of the left hand side (LHS) panel of fig.~\ref{NN_0} 
represent the configurations in the  
pre-training set, namely the three ground state configurations of $3$-states Potts model on 
a $L$ by $L$ lattice. In particular, each matrix entry stands for the Potts variable at the corresponding 
lattice site. Finally it should be pointed out that the output layer in our NN, which will 
be introduced later, is activated by a softmax function. As a result, the summation of
all components of any output $Q$-vector is 1.

One may wonder whether the NN with such purposely designed pre-training sets and labels
is capable of investigating
the phase transitions of the considered models efficiently. In the following, we will argue 
theoretically that the NN with the pre-training sets and labels introduced in the previous 
paragraph can be used as a valid and effective tool for uncovering the physics of the 
considered phase transitions. 

As described previously that for any $Q$-states Potts model with a given 
positive integer $Q$, the considered pre-training set consists of $Q$ objects
which are the associated ground state configurations. Furthermore, for the
$n$th object in the pre-training set, all its Potts variables take the same
positive integer value $n$. The labels for the objects in the pre-training set 
are $Q$-vectors and the correspondence between a training 
configuration and its label, i.e. a $Q$-vector, has already been defined 
earlier and can
be understood pictorially by fig \ref{NN_0}. Notice as one approaches 
the critical temperature $T_c$ from very low temperatures, any Potts variable 
of a configuration 
obtained by the Monte Carlo simulations may begin to take a different value 
from those of other Potts variables of that 
configuration. The closer to the $T_c$, the distribution of the values of the Potts variables 
are more random. For configurations determined above $T_c$, the Potts variables take their possible results 
in a totally random manner over the whole lattices.

The scenario of how the distribution of the values 
of the Potts variables changes from low temperatures to high temperatures will reflect in the NN output 
$Q$-vectors as follows. When one supplies the already trained NN with a configuration obtained at 
a temperature extremely far below $T_c$, one component (could be any of the $Q$ components)
of the output $Q$-vector has a numerical value much larger than those of other components
and the difference could be several order in magnitude. 
Such a result described above occurs because most of the Potts variables of the configuration take the same positive integer 
value. In addition, if one (testing) configuration is determined with a relatively lower 
temperature than $T_c$, besides the fact that the majority of the Potts variables take one particular 
$i$ in $\{1,2,3,...,Q\}$ as their value, theoretically the remaining part of the Potts variables will result 
in any allowed positive integer other than $i$ with a more or less equal amount of each. 
Consequently, the resulting output $Q$-vector has the 
following characteristics. Specifically, in addition to the fact that one component has the 
largest number $d$, other components of that $Q$-vector will begin to take 
non-negligible values, 
although these values may be relatively smaller than $d$.  
Finally if the values of the Potts variables of a testing configuration are randomly 
distributed all over the 
sites, then to the trained NN every one of the $Q$ ground states is the equally possible candidate 
for the testing configuration. For such a case, which corresponds to the situation that the simulated temperature $T$ is
relative higher than $T_c$, each component of the resulting $Q$-vectors     
has a value around $1/Q$. Based on these observations described above, as one moves from temperatures below $T_c$ to 
those above $T_c$, every component of the related output $Q$-vectors is approaching the same value $1/Q$.

As we have already described earlier, for testing configurations determined with the same box size $L$
and the same temperature $T$ (which is relatively lower than $T_c$), although the output $Q$-vector
may be different component-wisely from one configuration to another configuration,
for majority of these testing configurations the norms of the related output 
$Q$-vectors should be approximately the same ideally. 
This is because every possible integer for the Potts variables  
has equal footing with each other. As a result, the norm $R$ of the output 
$Q$-vectors will be a suitable (and better) quantity for our investigation.
For testing set determined by Monte Carlo
simulations with extremely low temperatures, the corresponding outcomes of $R$ are 1.
As one moves from low temperatures toward the vicinity of $T_c$, the norms of the NN 
output $Q$-vectors change from 1 to around $1/\sqrt{Q}$. For $T > T_c$, the resulting $R$
will be $1/\sqrt{Q}$ theoretically. In other words, the quantity $R$, 
namely the norm
of the NN output $Q$-vectors can be used as an effective observable to study the phase 
transitions of the considered $Q$-states Potts models. Furthermore, as we will demonstrate later,
$R$ is not only a valid quantity for estimating the critical temperatures, 
but is also an effective observable of determining the nature, namely whether they are
first order or second order, of the considered phase transitions. 

\subsubsection{The construction of NN}

After detailing the ideas behind the training sets and the corresponding 
labels used here, in the following we will introduce the NN considered 
in our study. In particular, we want to build a simple (deep learning) NN that 
enables us to carry out the related investigation. Notice the Keras library \cite{kera} 
is employed in the NN construction. In addition, the set up of the NN in this 
study consists of one input layer, one convolutional layer with kernel size 3 by 3, 
one average pooling layer with kernel size 2 by 2, and one output layer.
The initial matrix elements (weights) $W_{ij}$ connecting to the output layer, 
as well as the pooling kernel are the default ones of Keras.
 
Notice that in our original proposition, for a given $Q$-states Potts model
the training set, or more accurately the pre-training set contains $Q$ types 
of ground state configurations which may be not sufficient for a successful
training. Indeed, in a typical study of a phase transition using the NN
methods, depending on the batch size used in the training process, 
several hundred to (a) few thousand training configurations may be needed
in order to ensure the success of the training. One of the reasons of
considering sufficiently many training objects is to make sure that the
trained NN is capable of classifying any validation and testing configurations
with high precision. Furthermore, during the minimization procedures, the 
strategy of using enough amount of training objects also helps in reaching 
good results for the desired weights connecting different layers (Of course,
one has to pay attention to some subtle issues such as overfitting as well).
Therefore, in our investigation multiple copies of
the original $Q$ types ground state configurations are used as the actual
training set. This is the reason why the $Q$ types ground state configurations
are called the pre-training set in this study. Notice in our NN procedures, 
after the step of one-hot encoding, a random shuffle among the training 
objects is performed (Such an operation can be one of the standard steps for 
the training of a NN). As a result, the training objects contained in  
any batch may be significantly different from those of other batches. In addition, 
with the strategy of carrying out a random shuffle on the training objects, 
if one performs two training processes, then the initial objects in each (ordered) 
batch of these two training processes will be different as well. Therefore, 
technically speaking, while originally one has $Q$ configurations in the 
pre-training set, the use of multiple copies of the $Q$ types ground state 
configurations as the actual objects in the training set does practically 
enlarge the space of the pre-training set. 

We find that when $N \ge (n \times {\text{batch size}})$ 
copies of the ground state configurations are used as the 
training set, the obtained NN results are stable. Here the suitable value of 
$n$, which is a positive integer, requires some trial investigation. In this study the outcomes determined with 
$N = 200$ (and the used batch size is 20) are shown explicitly. It should be pointed out as 
well that the ratio between $N$ and the used batch size cannot be too large. This is 
because when this ratio is too large, the minimization procedures will be unstable due to the facts 
that each of many batches may contain the same type of training objects in itself, 
or numerous batches could have almost-identical (or identical) training items among themselves.
Finally we would like to emphasize the
fact again that in our calculations, 
a configuration on a $L$ by $L$ lattice is represented by a $L$ by $L$ matrix. 
Furthermore, the matrix elements are the values of the Potts variables located 
at the associated sites. The details of the training for our constructed NN for 
any $Q$-states Potts model is as follows.  
     
\begin{figure}
%\vskip0.5cm
\begin{center}
\includegraphics[width=0.5\textwidth]{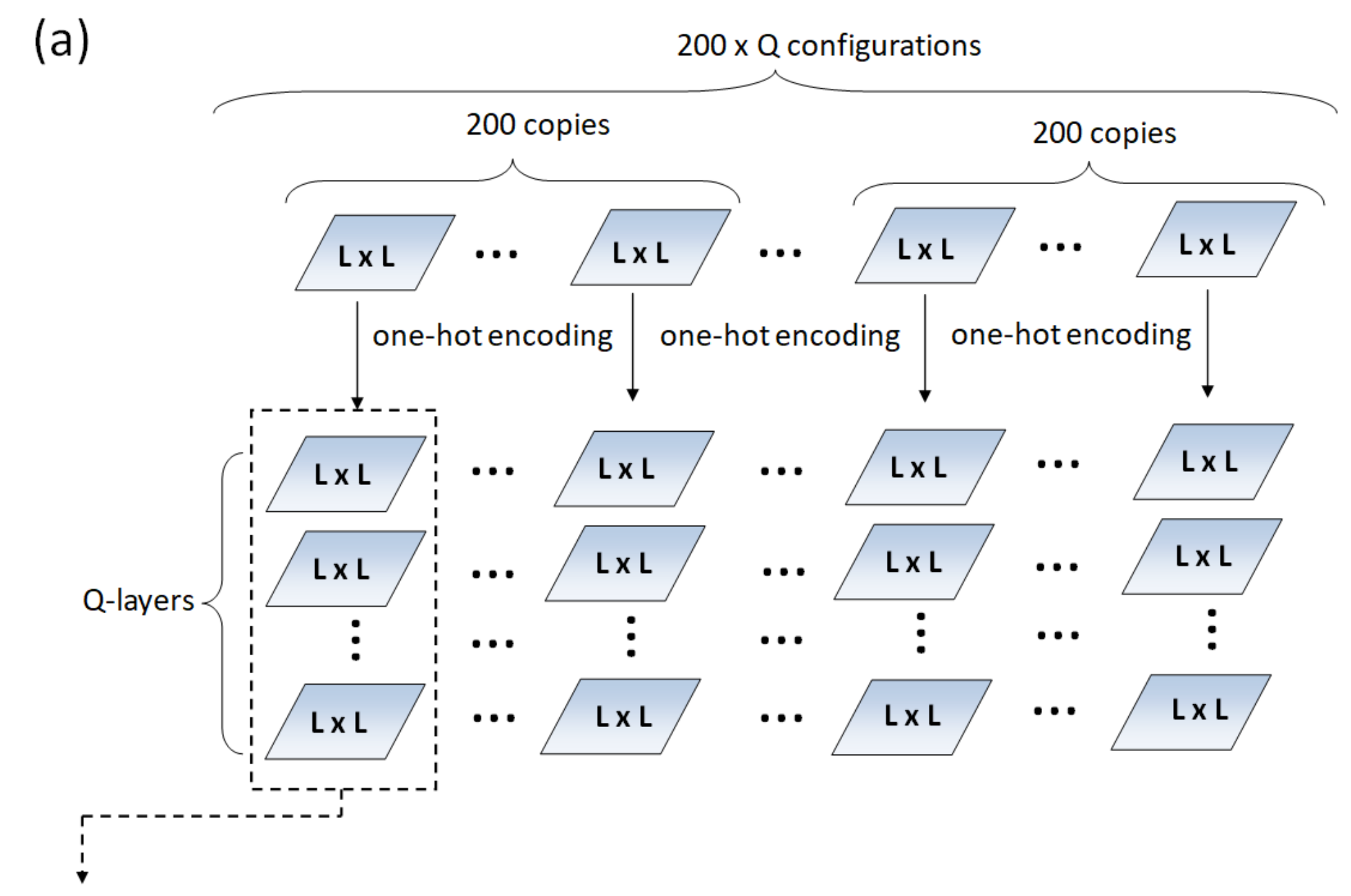}
\includegraphics[width=0.5\textwidth]{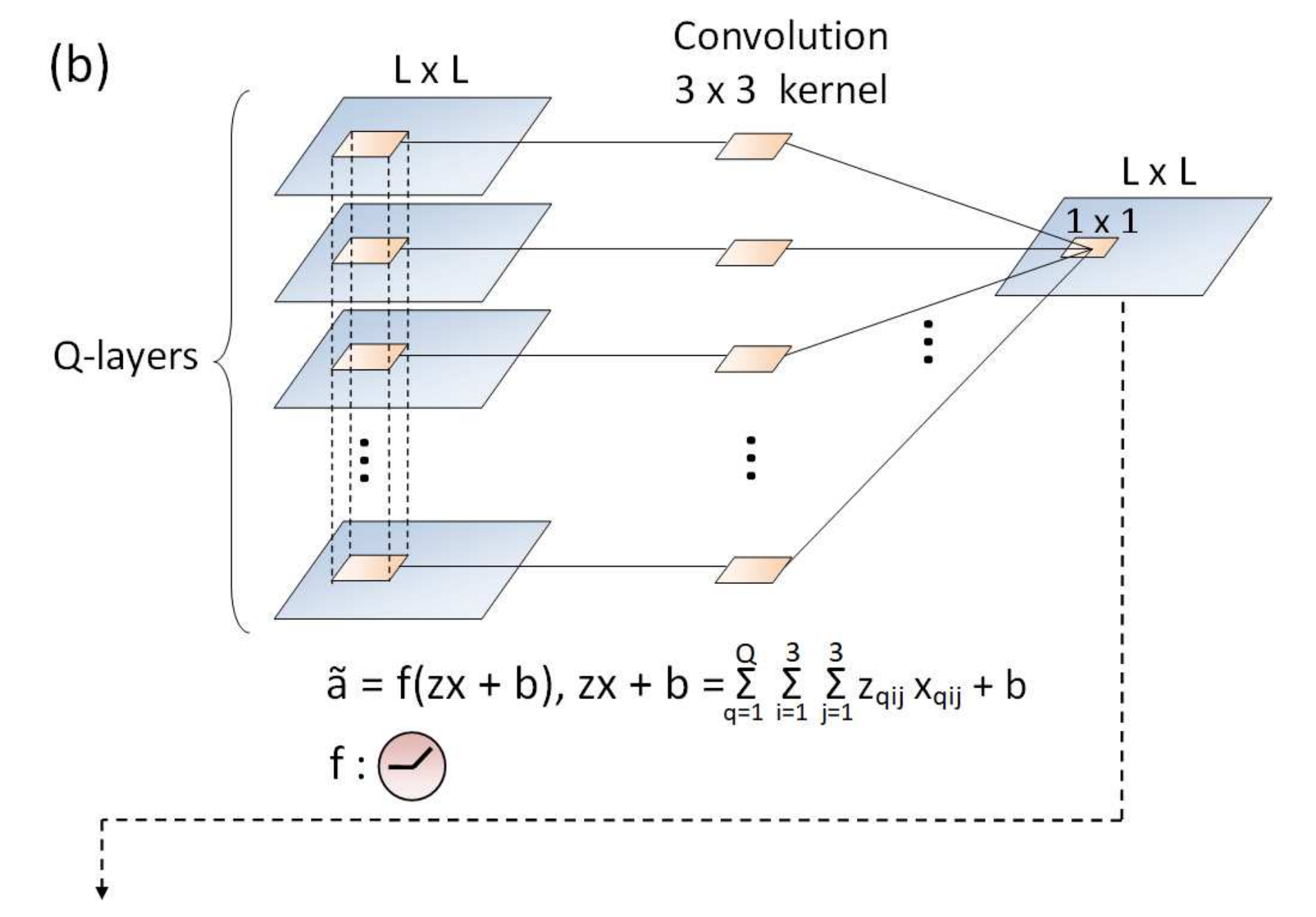}
\includegraphics[width=0.5\textwidth]{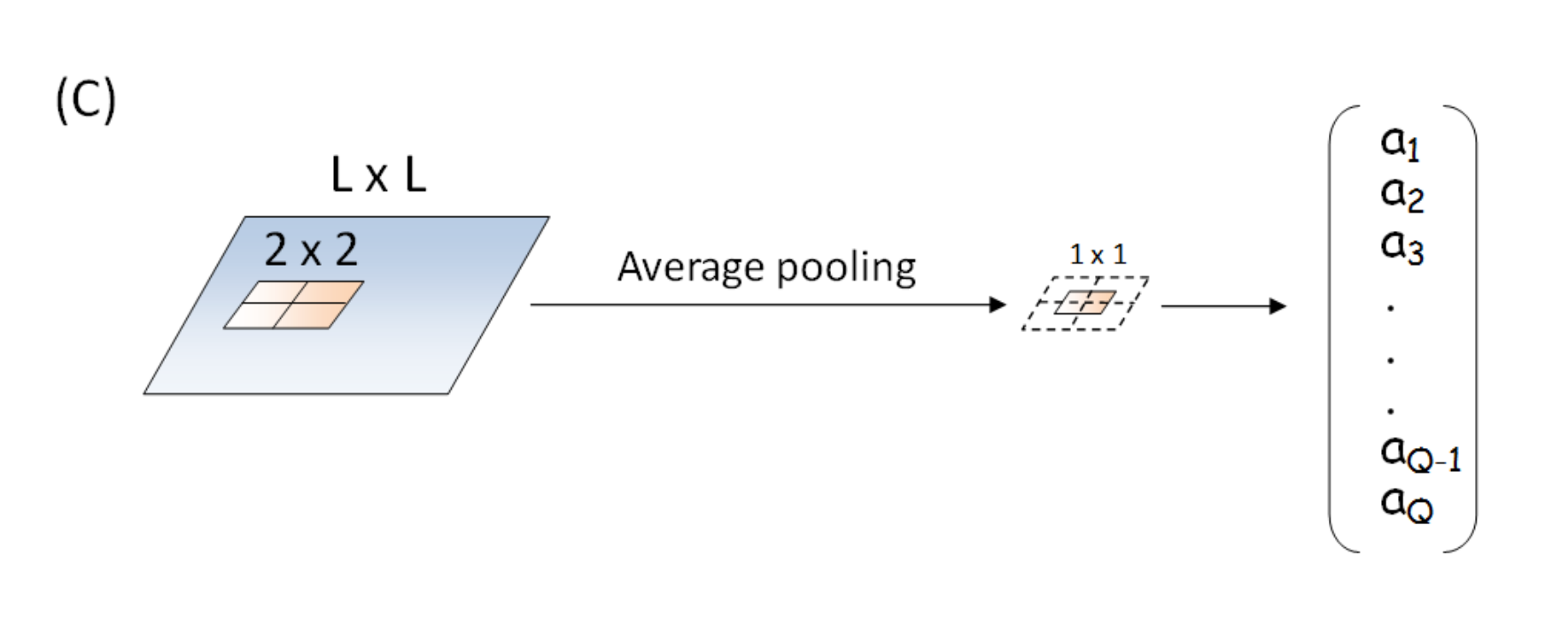}
\end{center}\vskip-0.7cm
%\vskip2.0cm
\caption{The set up NN in this study. The weights connecting different layers
are not shown explicitly. 
While panel (a) demonstrates the preparation of training set, panels (b)
and (c) show how the convolution and average pooling work on one of
the $200 \times Q$ training objects, respectively. In the actual calculations,
results from all the $200 \times Q$ training objects are taken into account. 
Notice after the step of one-hot encoding, a random shuffle among
the 200 $\times Q$ objects is conducted.}
\label{NN_1}
\end{figure}

For a given fixed box size $L$
and for each number $i$ in $\{1,2,3,...,Q\}$, 200 identical
$L\times L$-matrices are created.
In particular, every element in these matrices
has an initial numerical value $i$. This procedure leads
to $200 \times Q$ objects.
These $200 \times Q$ objects are then
one-hot encoded, resulting in yet another $200 \times Q$ objects for which each of them consists of
$Q$ layers of $L \times L$-matrices. These $200 \times Q$ objects are the training set 
for the set up NN. The procedure of creating the training sets from the 
pre-training configurations is depicted in the sub-fig.~$(a)$ of fig.~\ref{NN_1}.

The convolutional layer is activated by rectified linear
functions (ReLUs), see the $f$ at the bottom of sub-fig.~$(b)$ of
fig.~\ref{NN_1}.
Notice in the sub-fig.~$(b)$ of fig.~\ref{NN_1}, $\tilde{a}$ are the
outcomes
of applying the ReLUs to the results after the convolutional layer.
We have considered different initial filter
kernels, and find that there are some restrictions on the corresponding kernel
matrix elements. Specifically, each matrix element should not be too large,
otherwise one hardly obtains satisfactory results.
In addition to the input layer, the convolutional layer, and the output layer,
the built NN contains an
average pooling layer as well (sub-fig.~$(c)$ of fig.~\ref{NN_1}). The full pictorial representation of the set
up NN is depicted in fig.~\ref{NN_1}.

The main algorithm considered in the built NN is the mini-batch. As a result,
the $200 \times Q$ objects of the training sets will be permuted randomly before 
the actual training is conducted. Furthermore,
we exploit the Adam optimizer of Keras to update the weights of the constructed
NN. For the loss functions we use "categorical crossentropy" which is defined as
\begin{equation}
C = -\frac{1}{n} \sum_{x}\sum_{j}^{Q}\left[y_j \ln a_j + (1-y_j)\ln (1-a_j)\right],
\end{equation}
where $n$ is the total number of objects in the training set,
$a_j$ are the outcomes obtained after applying all the constructed layers.
In addition, $x$ and $y$ are training inputs and
the corresponding designed outputs, respectively. Finally, in our study
we use the default $\eta$ as the learning rate.

In our investigation, as we have already mentioned earlier that various batch 
sizes, ranging from 10 to 80, are used to examine the stability of the output results 
with respect to the amount of copies of the pre-training set used as the objects in the 
training set.  
In addition, the number of epoch is
confined by an early stop criterion and $L2$-regularization is imposed in
calculating the loss functions. The output layer in the set up NN is activated by a softmax
function. As a result, the sum of all components of a output vector is 1.
Notice there are several tunable parameters in our NN and we will make a remark regarding
this later.

\subsubsection{The validation and testing of the constructed NN}
When the training of a built NN is done, one has to carry out the validation process
to ensure the success of the training, namely to confirm the trained NN is capable
of recognize every input from the validation set with high accuracy.   
For $Q$-states Potts models, we use 100 copies of the corresponding $Q$ ground state 
configurations as the validation sets. Furthermore, we find that the accuracy of the validation for
the NN used here is 100 percent.
It should be pointed out that while a random permutation is performed among the objects in the training sets
before the actual training is executed, 
the training and validation sets considered in our study are actually the same from a general point of view.
As a result, the validation conducted here should be viewed as a guidance for the success of 
the training stage. Whether the constructed NN can fulfill the desired tasks of determining 
the critical points as well as uncovering the nature of the studied phase transitions, 
should be evaluated based on if the NN results are consistent with the 
established outcomes obtained by the traditional methods.        
  
For each temperature, the testing set for the constructed NN consists of 
several thousand configurations generated by MC simulations (separating
from each other by few thousand updates).
Notice with the built NN, as we already explained earlier, 
at the temperatures (far) below $T_c$, ideally one will find that 
each component of the output vector for every configuration generated by 
Monte Carlo simulations is zero, except one component which takes 
the value of 1. The norms of such vectors are $1$.
On the other hand, for the temperatures above $T_c$, 
since the values of Potts variables are in a totally random manner over 
all the lattice sites, the trained NN is not able to definitely decide which 
of the $Q$ possible (ground) states the system belongs to. 
Notice for such cases, every Potts variable
will take any of the integers in $\{1,2,,..,Q\}$ with equal probabilities. 
Therefore it is expected that each component of the corresponding output 
vectors has $1/Q$ as its value theoretically 
(The norms of these vectors are $1/\sqrt{Q}$). 
Based on these observations, similar to the conventional approaches, 
investigating the norms and their histograms of the output $Q$-vectors may
shed some light on locating the critical temperatures as well as
uncovering the nature of these phase transitions. In particular, for any given $Q$-states
Potts model, the associated
critical temperature $T_c$ should lie within the temperature 
interval $\left(T_1,T_2\right)$, where $T_1$ and $T_2$ are the temperatures 
at which $R$ begins to drop from its upper 
bound 1 and reaches its lower bound $1/\sqrt{Q}$ for the first time, respectively.

\begin{figure}
%\vskip0.5cm
\begin{center}
\vbox{
\includegraphics[width=0.38\textwidth]{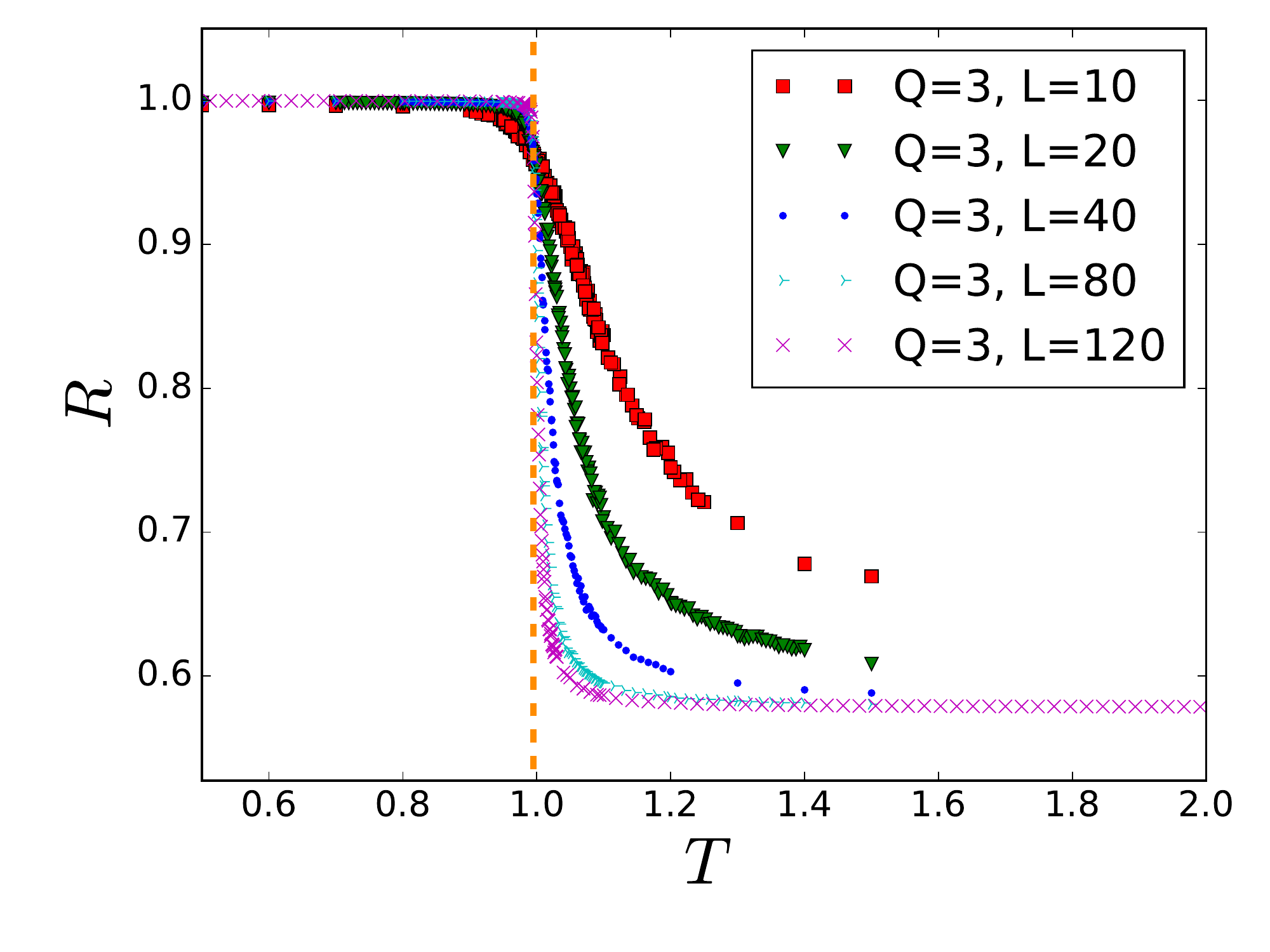}
\includegraphics[width=0.38\textwidth]{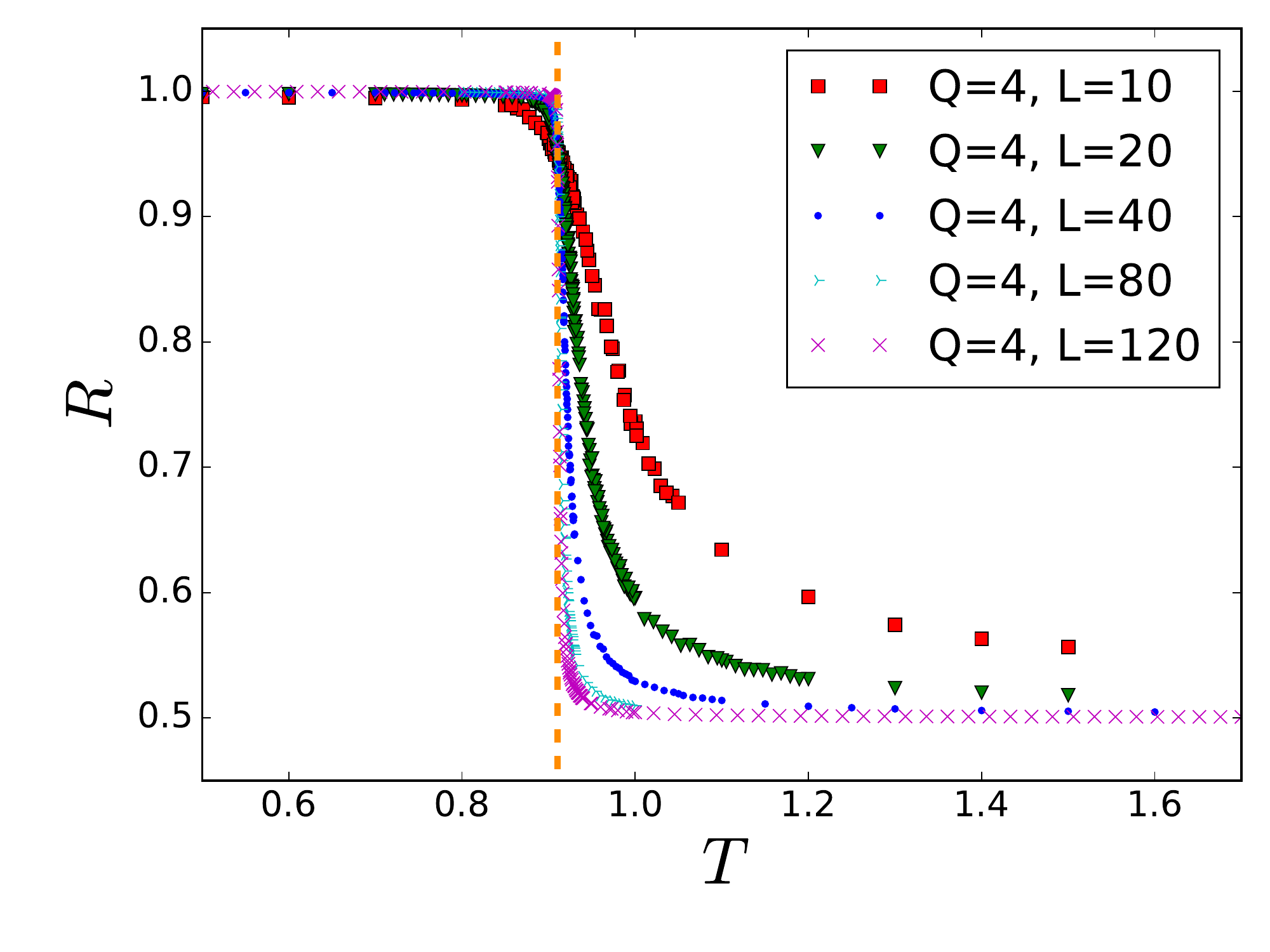}
\includegraphics[width=0.38\textwidth]{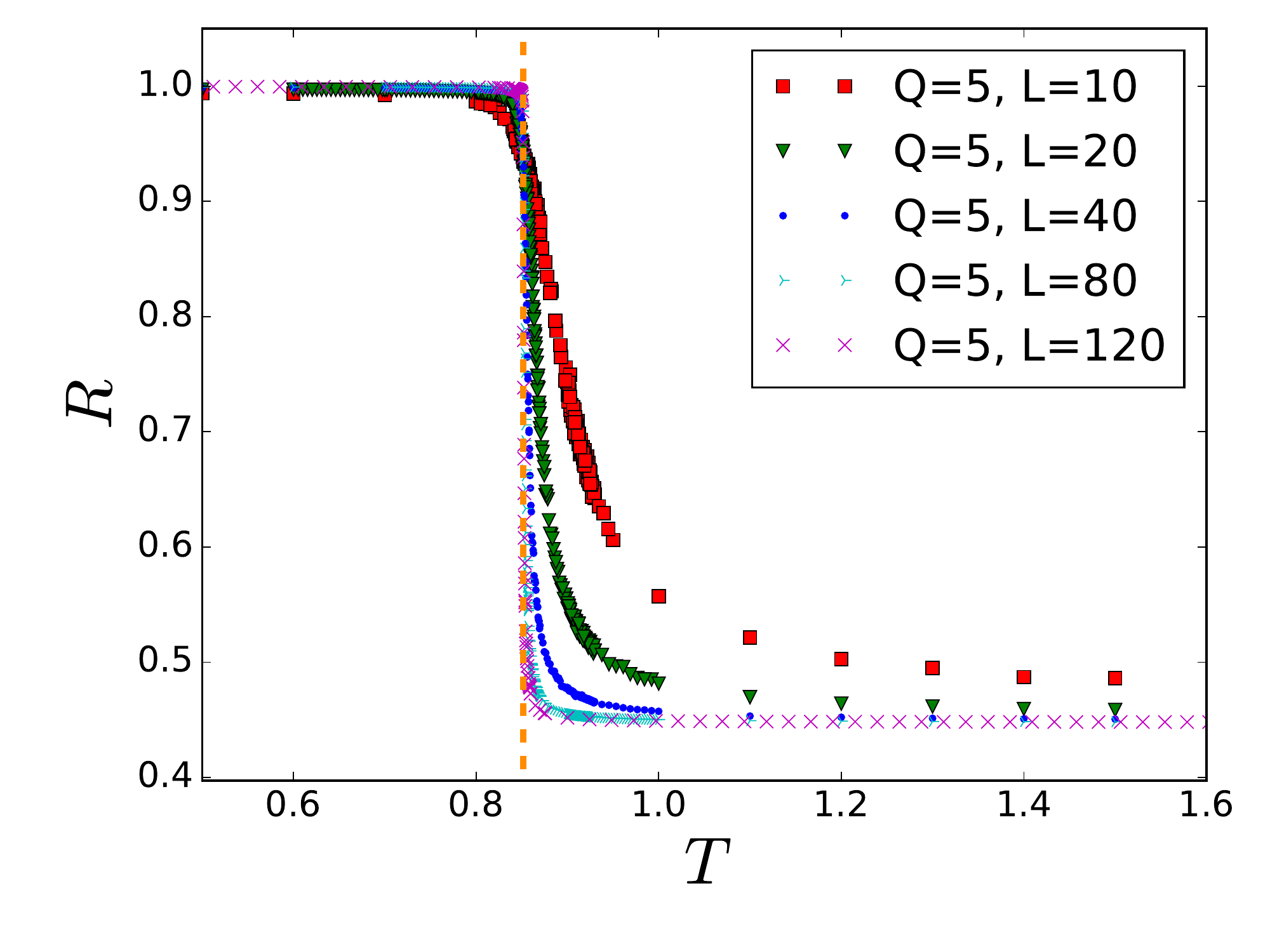}
\includegraphics[width=0.38\textwidth]{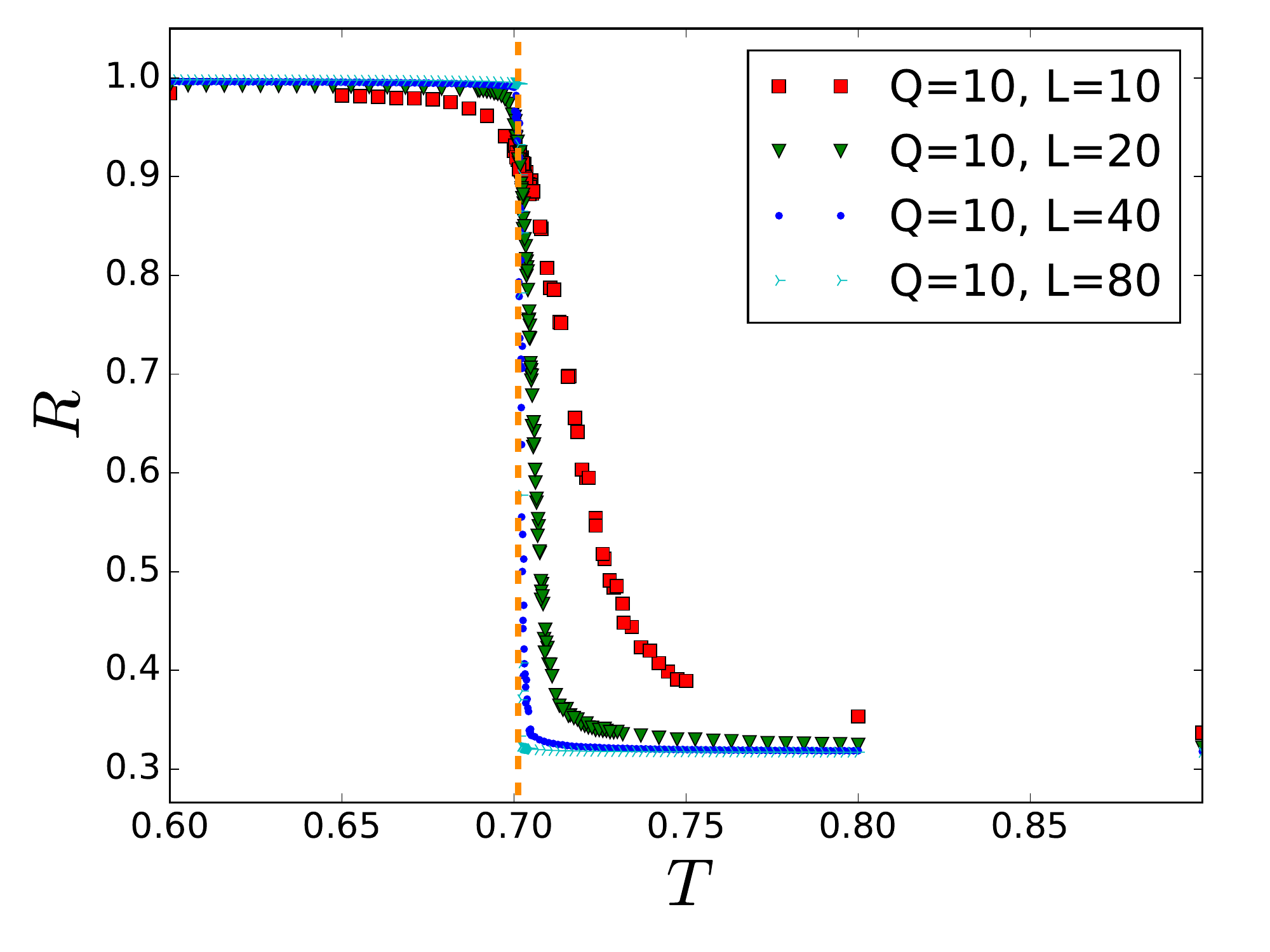}
}
\end{center}\vskip-0.7cm
%\vskip2.0cm
\caption{ (From top to bottom) The norms $R$ of the NN output vectors as functions of
$T$ for 3-, 4-, 5-, and 10-states Potts models (and for several linear lattice sizes $L$). 
For comparison (and better visualization as well), in each panel a vertical dashed line 
which intersects the $T$-axis at the known $T_c$ in the literature is added by hand.
}
\label{T_c}
\end{figure}

Remarkably, even with the simple training sets (and labels) considered here, 
the constructed NN can be used to accurately estimate the critical temperatures
$T_c$ of the investigated phase transitions. The norms $R$ of the NN output vectors 
as functions of $T$ for 2D 3-, 4-, 5- and 10-states Potts models are shown
in fig.~\ref{T_c}. We find that for each of the studied models, as $L$ increases, 
the range of $T$ for which $R$ drops from 1 to $1/\sqrt{Q}$
converges quickly to a narrow region. In particular, the known critical temperatures
in the literature for the investigated phase transitions 
lie within these regions of $T$. For example, the $T_c$ for 3-states Potts 
model is given by $T_c\sim 0.99497$ in the literature, and one clearly 
sees from the most top panel of fig.~\ref{T_c} that as the box size $L$ 
increases, the range of temperatures where
$R$ drops from 1 and saturates to $1/\sqrt{3}$ converges to a narrow 
region containing the expected $T_c \sim 0.99497$.  

\subsubsection{The nature of the considered phase transitions}
Notice while it is beyond doubt that the built NN in this study can precisely 
determine the $T_c$ of the considered phase transitions, fig.~\ref{T_c} does not
provide us with sufficient information to decide whether they are
first order or second order. To uncover the nature of these phase
transitions of 2D Potts models using the constructed NN shown in 
fig.~\ref{NN_1}, we turn to studying the histograms of $R$. 
This strategy is motivated by one of the traditional methods, as shown 
previously in the subsection of Monte Carlo results. Notice since for 
temperatures which are relatively far below and above $T_c$, 
the numerical values of $R$ are given by 1 and $1/\sqrt{Q}$, respectively, 
in the histogram(s) of $R$ one may observe two peaks located at $R = 1$
and $R = 1/\sqrt{Q}$ near $T_c$. In particular, for a first order phase transition
the relative heights between the peak(s) and the trough should increase 
with the linear box sizes $L$. Here we would like to emphasize again the fact that 
for a first order phase transition, the appearance 
of two peak phenomena in the histograms of certain observables is because the system is 
equally probable in either one of two possible states. For our case, one of these two probable 
states has $R=1$ and the other has $R=1/\sqrt{Q}$. Therefore, studying the histograms of $R$ 
may shed some light on uncovering the nature of the investigated phase transitions.  

\begin{figure}
%\vskip0.5cm
\begin{center}
\vbox{
\includegraphics[width=0.325\textwidth]{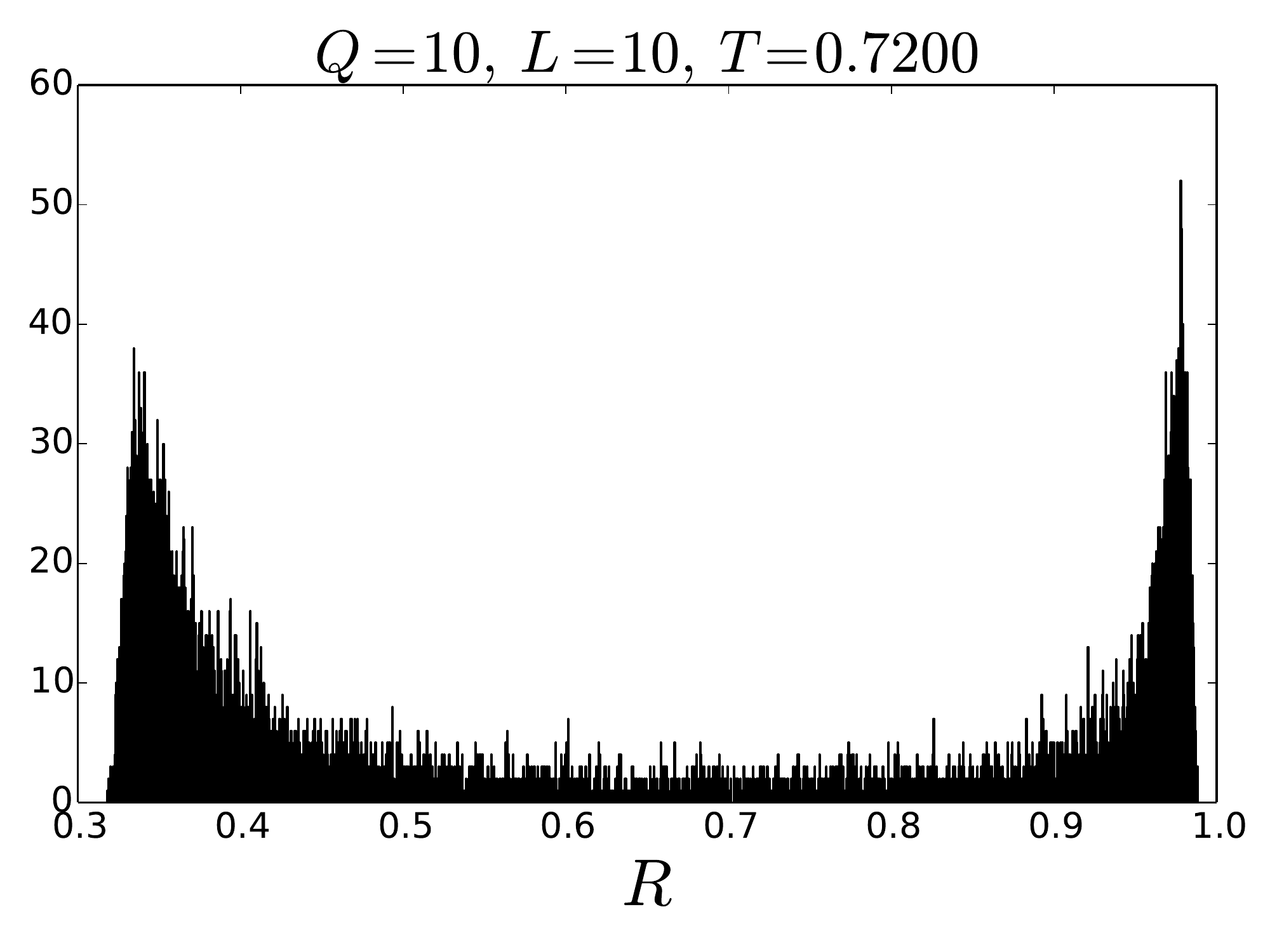}\vskip0.25cm
\includegraphics[width=0.325\textwidth]{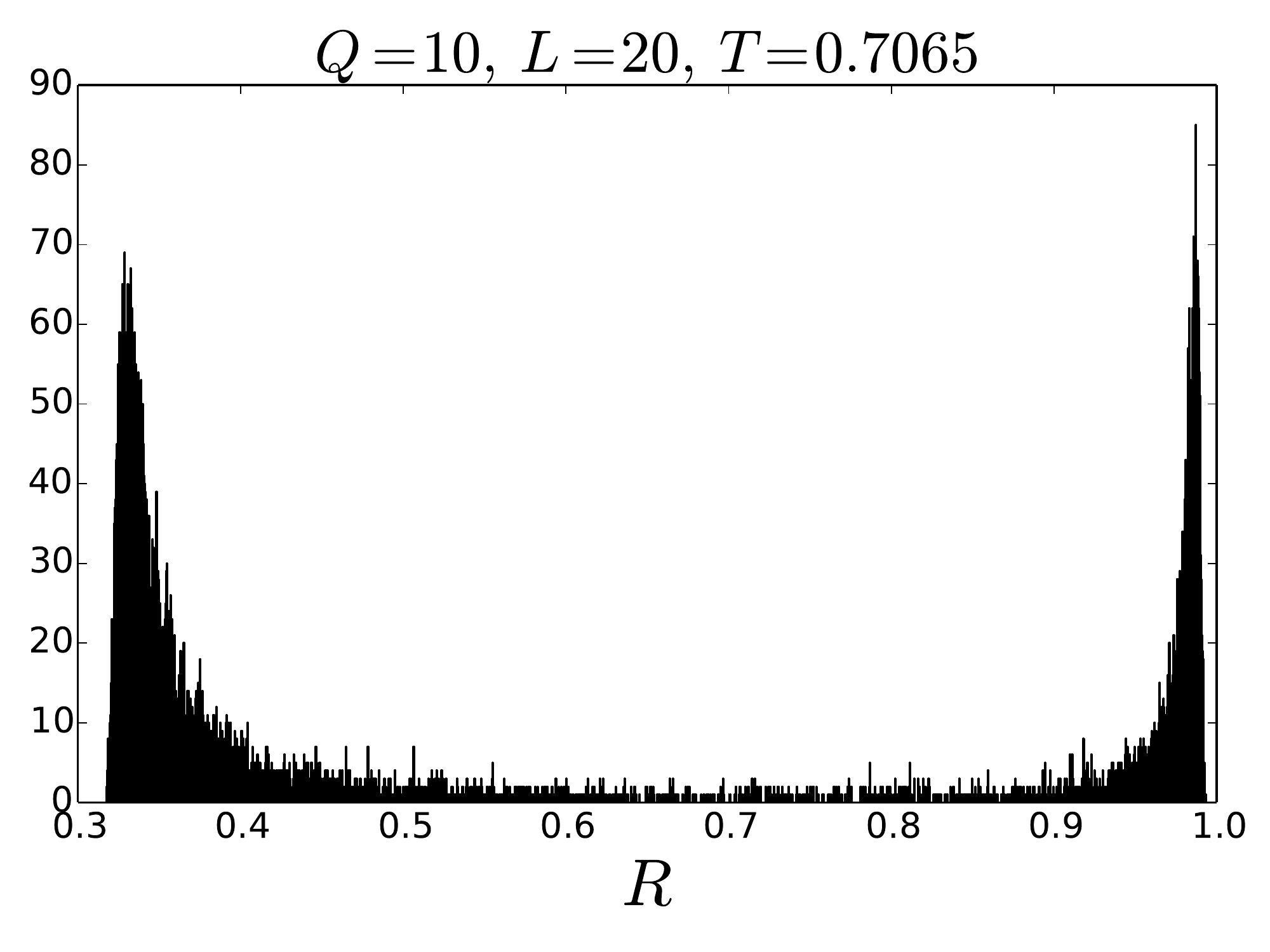}\vskip0.25cm
\includegraphics[width=0.325\textwidth]{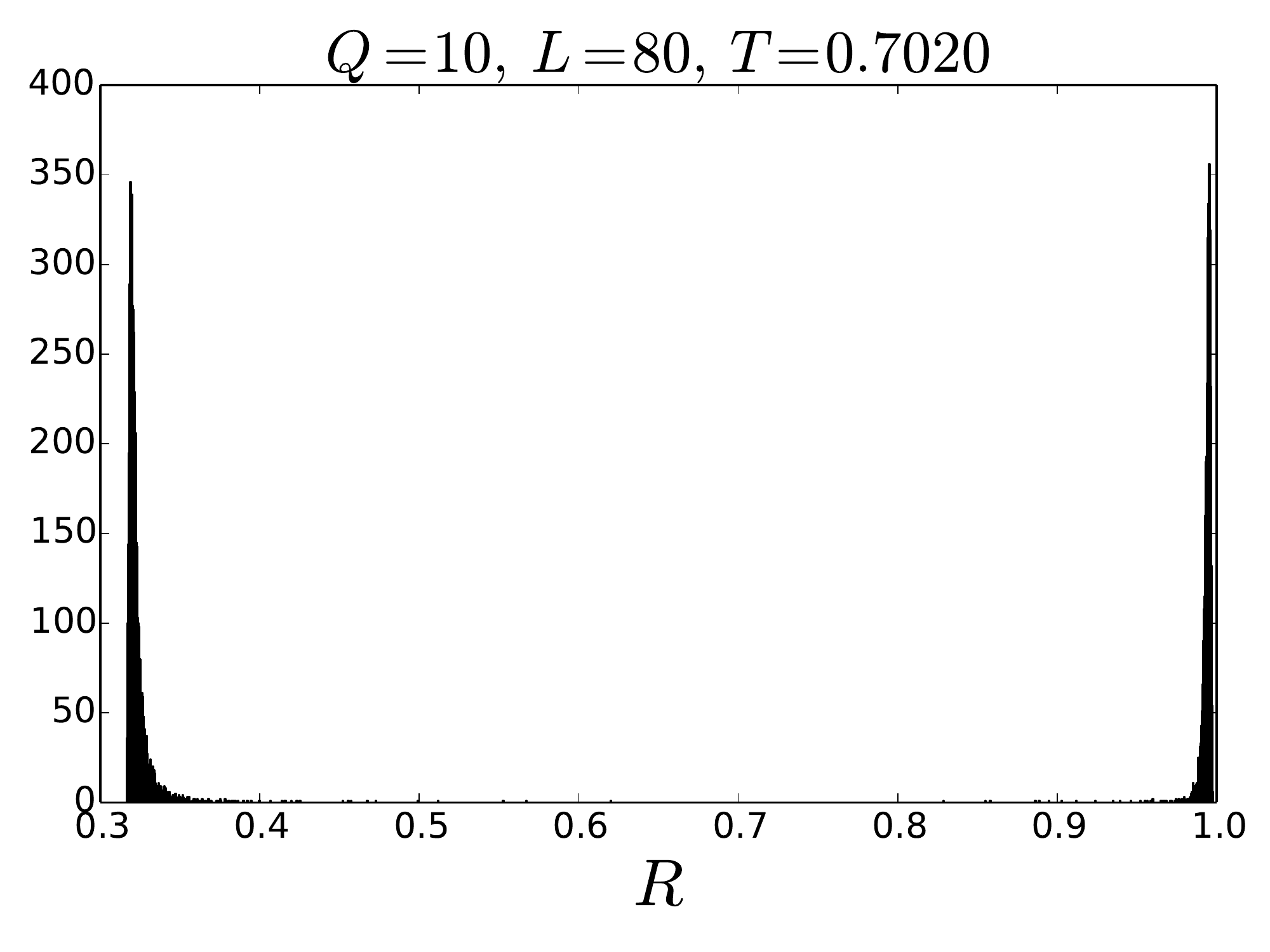}
}
\end{center}\vskip-0.7cm
\caption{Histograms of $R$ for $L = 10$ (top), 20 (middle),
and 80 (bottom) for 2D $10$-states Potts model. Around ten thousand data 
points close to the corresponding
$T_c$ are generated for each of the plots. Data are recorded once for every 
2000 updates after the thermalization. The histograms are produced by the 
``hist'' function of pylab \cite{pyl}.}
\label{NN_2}
\end{figure}

Remarkably, such a scenario introduced in the previous paragraph does happen in the histograms of $R$ for 2D $10$-states 
Potts model. The results depicted in fig.~\ref{NN_2} are
the histograms of $R$, obtained with $L = 10$, 20 and 80 close to their 
corresponding
critical temperatures ($T_c(L)$), for
2D 10-states Potts model. 
Fig.~\ref{NN_2} shows that two peaks phenomena appear for all three lattices, 
and this phenomena becomes more and more noticeable as one increases the linear 
lattice size $L$. In particular,
almost all the obtained results of $R$ for $L=80$
are located at $R=1.0$ and $R=1/\sqrt{10} \sim 0.31623$.    
This implies that the phase transition
of 2D $10$-states Potts model is first order as expected. Notice the 
temperatures at which this two peaks phenomenon is the most obvious may be 
slightly different from those shown in fig.~\ref{MC_2}. This small discrepancy 
can be attributed to the fact that $T_c(L)$ are observable dependence.

\begin{figure}
\vskip0.5cm
\begin{center}
\vbox{%\hskip-0.3cm
\includegraphics[width=0.325\textwidth]{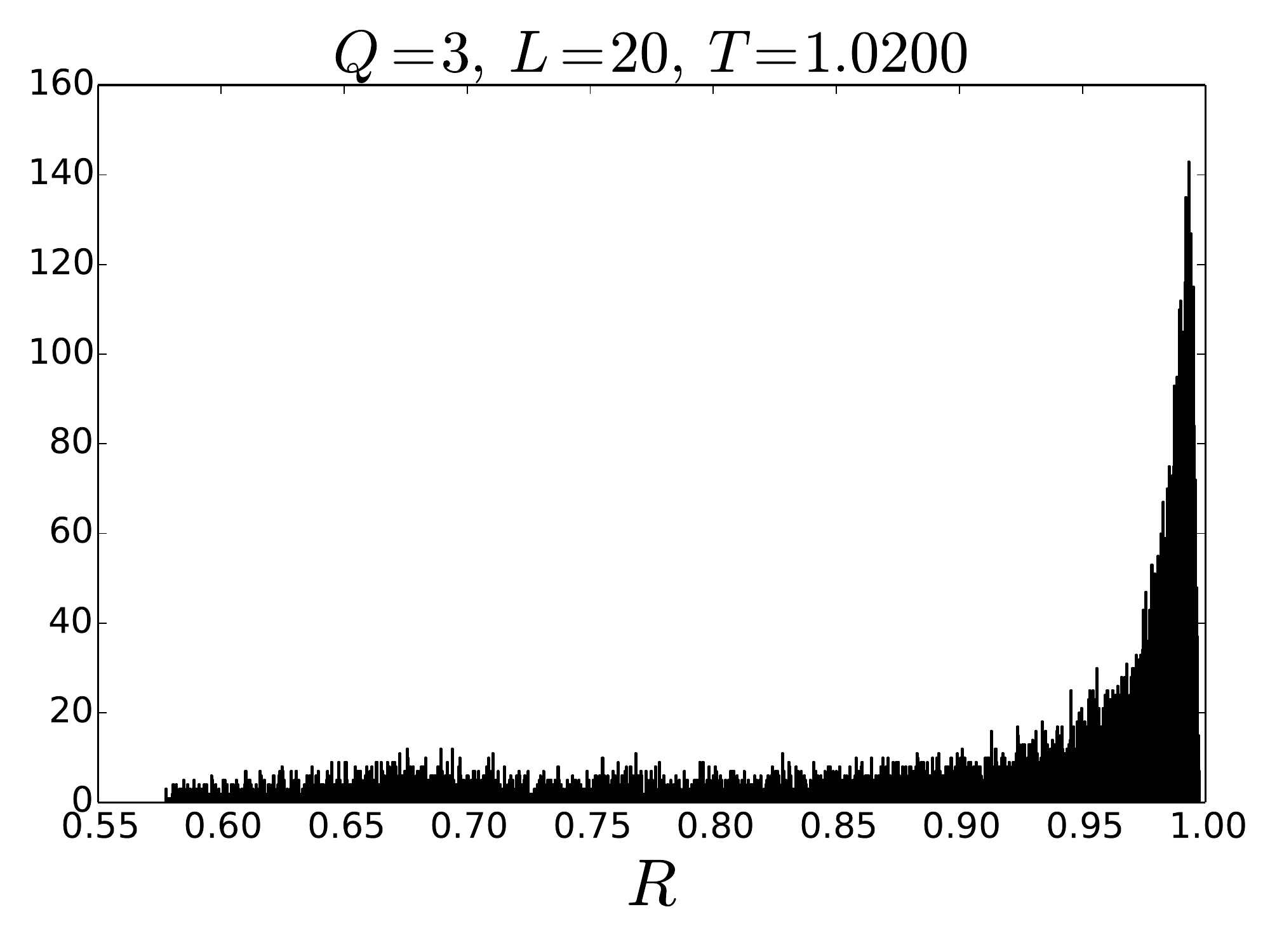}\hskip0.05cm
\includegraphics[width=0.325\textwidth]{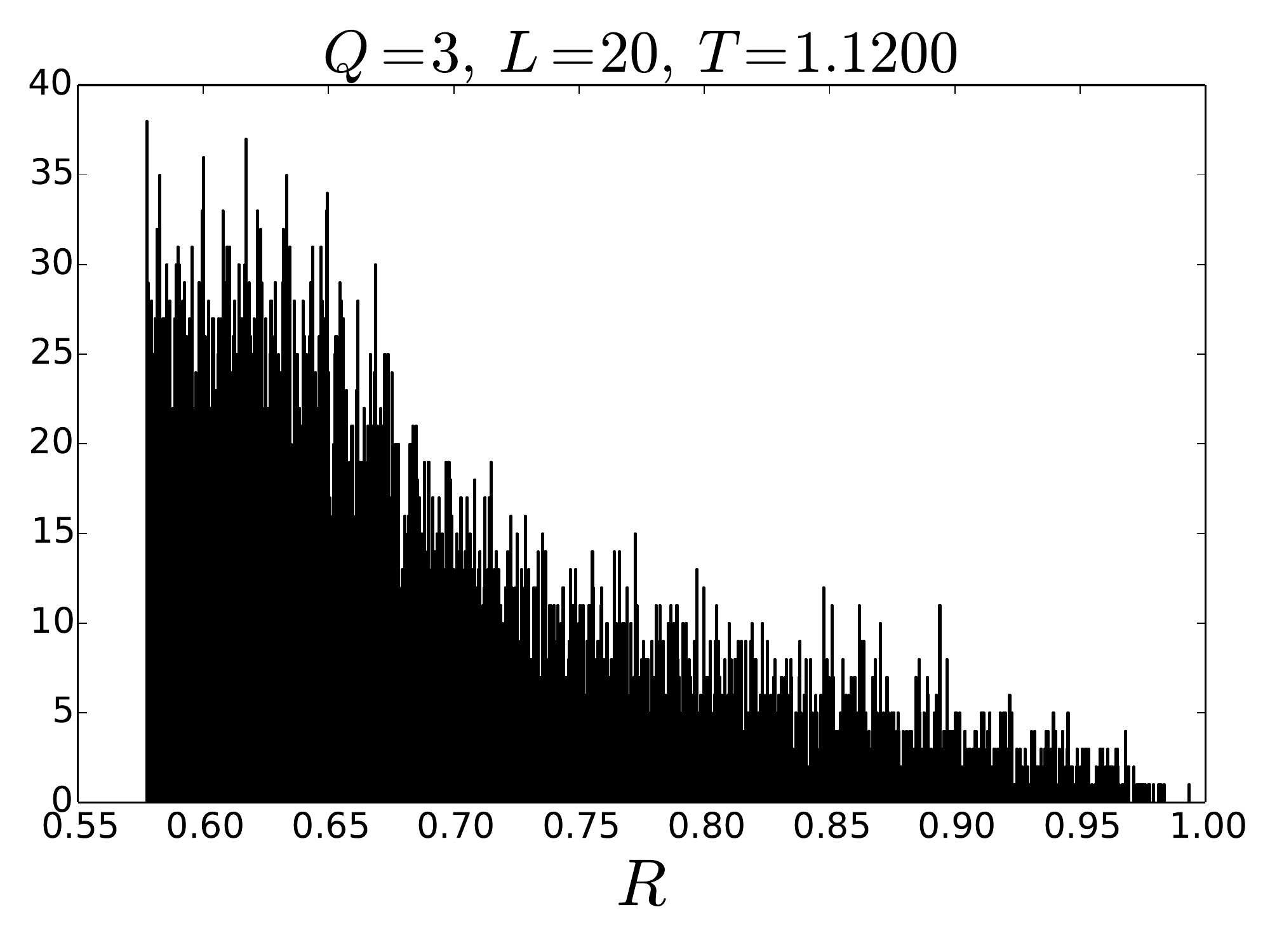}
}
\end{center}\vskip-0.7cm
\caption{Histograms of $R$ for $L = 20$ at temperatures $T=1.02$ (top) and $T=1.12$
(bottom) for 
2D $3$-states Potts model. Around ten thousand data are generated for each of 
the plots. Data are recorded once for every 
2000 updates after the thermalization. 
The histograms are produced by the ``hist'' function of pylab \cite{pyl}.}
\label{NN_3}
\end{figure}

\begin{figure}
\vskip0.5cm
\begin{center}
\vbox{%\hskip-0.3cm
\includegraphics[width=0.325\textwidth]{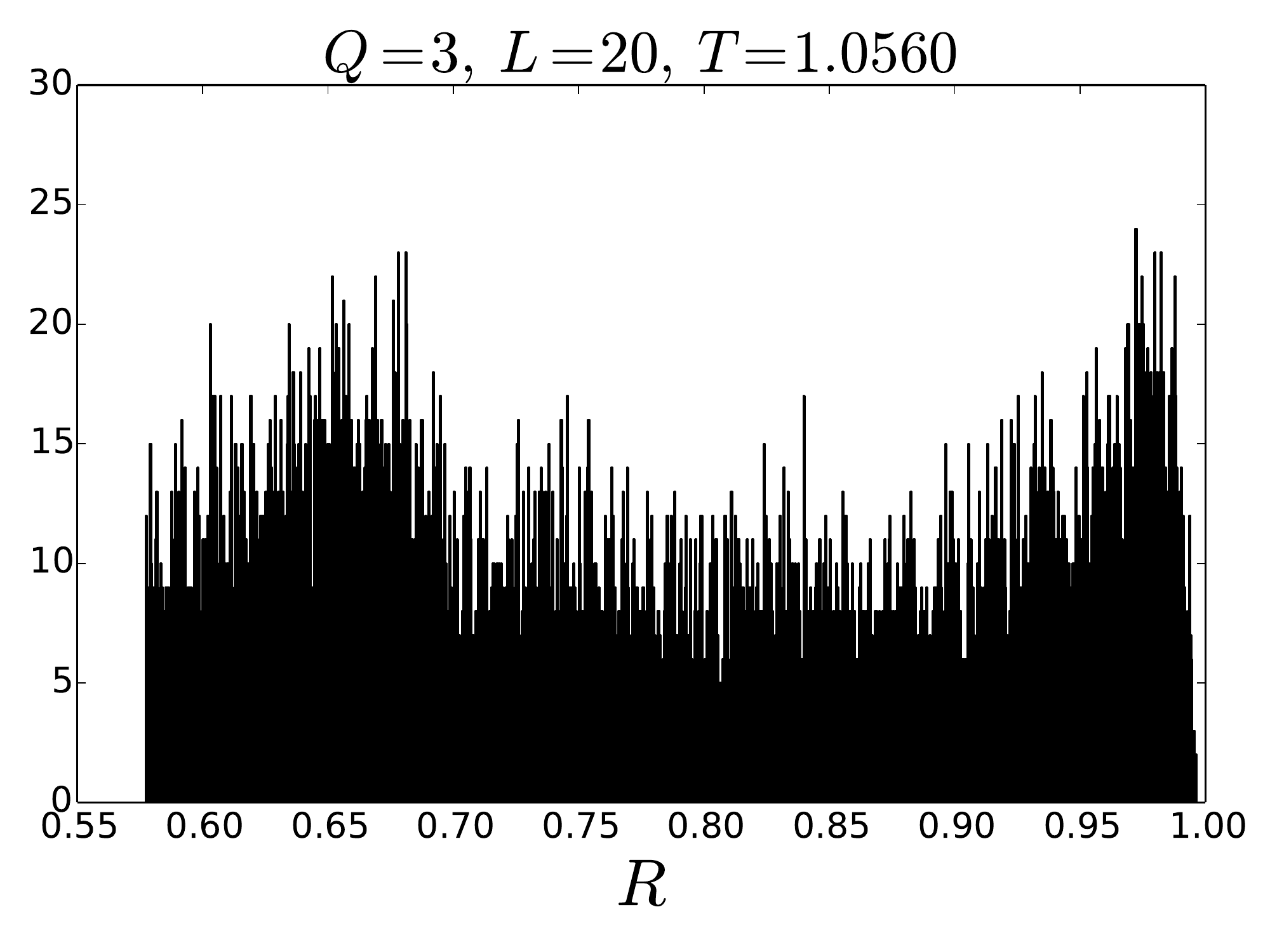}\hskip0.05cm
\includegraphics[width=0.325\textwidth]{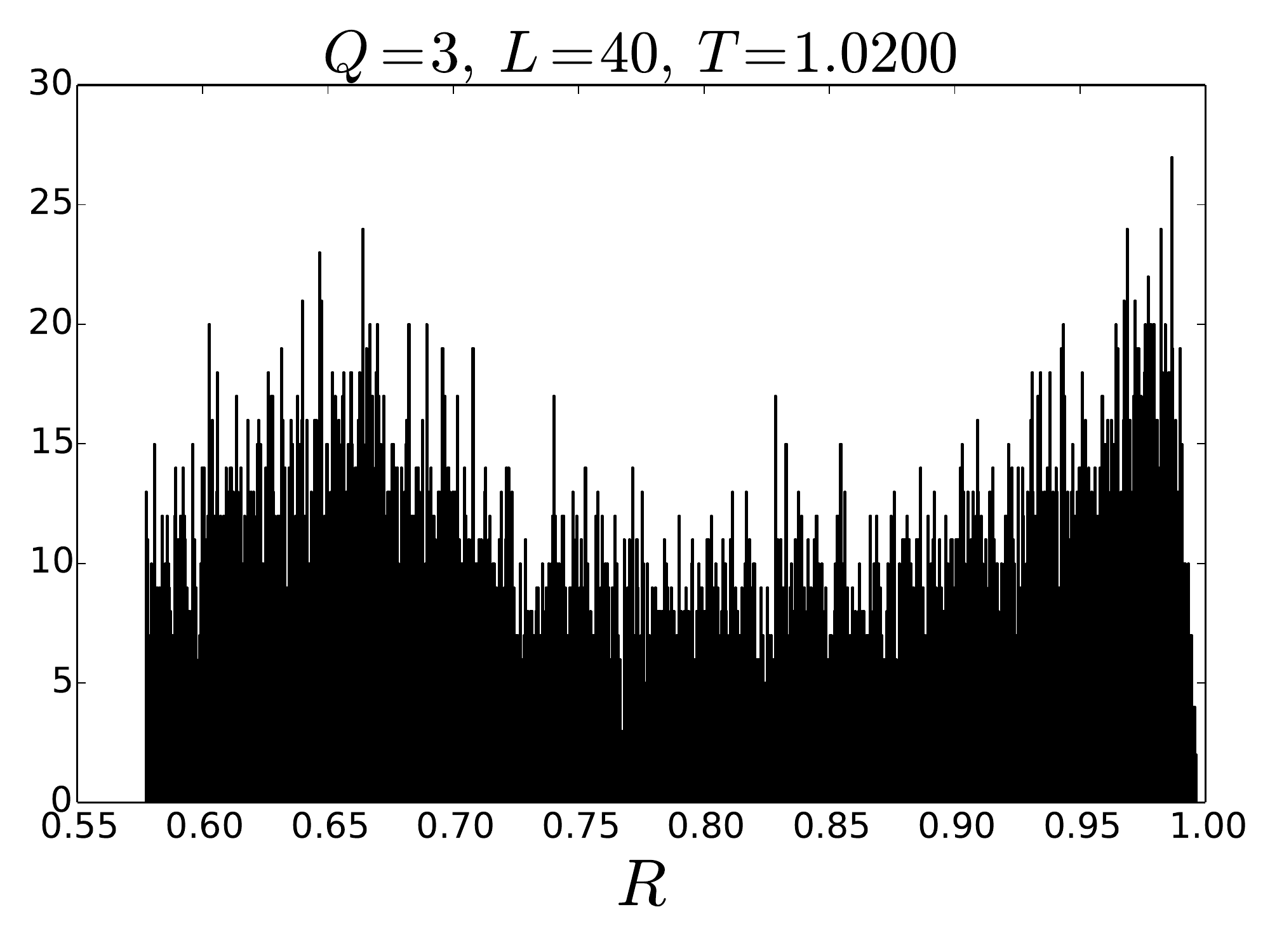}\hskip0.05cm
\includegraphics[width=0.325\textwidth]{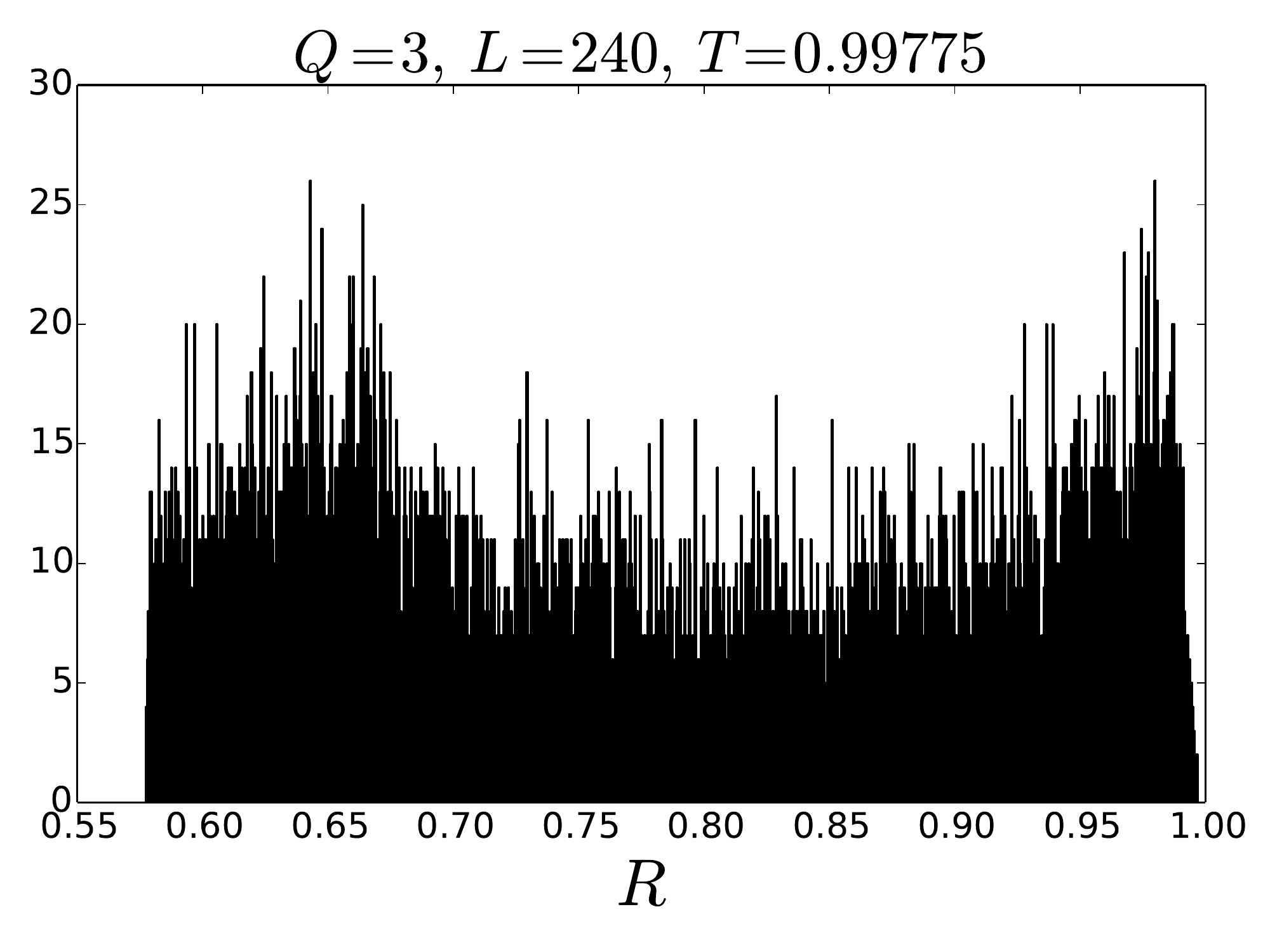}
}
\end{center}\vskip-0.7cm
\caption{Histograms of $R$ for $L = 20$ (top), 40 (middle), and 240 (bottom) 
for 
2D $3$-states Potts model. Around ten thousand data close to the corresponding 
$T_c$ are generated for each of the plots. Data are recorded once for every 
2000 updates after the thermalization. 
The histograms are produced by the ``hist'' function of pylab \cite{pyl}.}
\label{NN_4}
\end{figure}

Figures \ref{NN_3} and \ref{NN_4} show the
histograms of $R$ close to $T_c$ for 2D 3-states Potts models.
From these figures we find that as one moves from low temperatures to high 
temperatures, the histograms of $R$ for 2D 3-states Potts model evolve from 
a configuration 
with (only) 
one peak locating near $R=1$ to a configuration with (only) one peak near 
$R=\frac{1}{\sqrt{3}}$,
without two peaks structure like those of figure~\ref{NN_2} appearing for the 
intermediate temperatures. From this observation as well as the fact that in 
figure \ref{NN_4} the relative heights between the peaks and trough 
are almost the same for $L=20$, 40, and 240, one concludes that the 
corresponding phase transition is second order. 
These two results of 2D $3$- and $10$-states Potts models obtained using the 
constructed NN in this study agree with those determined by other 
traditional methods.
 
Notice for 10-states Potts model, the temperature at which the two peaks 
structure is the most obvious for $L=80$ is around 0.702 which agrees very well with
the expected $T_c \sim 0.70123$. In addition, for 2D 3-states Potts model, 
the temperature at which the histogram of $R$ for 
$L$ = 240 is the most flat is around $T \sim 0.99775$. 
The temperature $T\sim 0.99775$ again matches nicely with the 
theoretical prediction $T_c \sim 0.99497$. These results provide convincing evidence that
the NN constructed based on our ideas is able to locate the critical points of 
the considered models
in the relevant parameter spaces. Furthermore,
it is capable of detecting the nature of the studied phase transitions 
efficiently as well.   
Finally, by studying the 
histograms of $R$ for 2D $2$-,$4$-,and $5$-states Potts models, we also 
arrive at conclusions consistent with the known results in the literature 
regarding the nature of these phase transitions.

\begin{figure}
%\vskip0.5cm
\begin{center}
\vbox{
\includegraphics[width=0.325\textwidth]{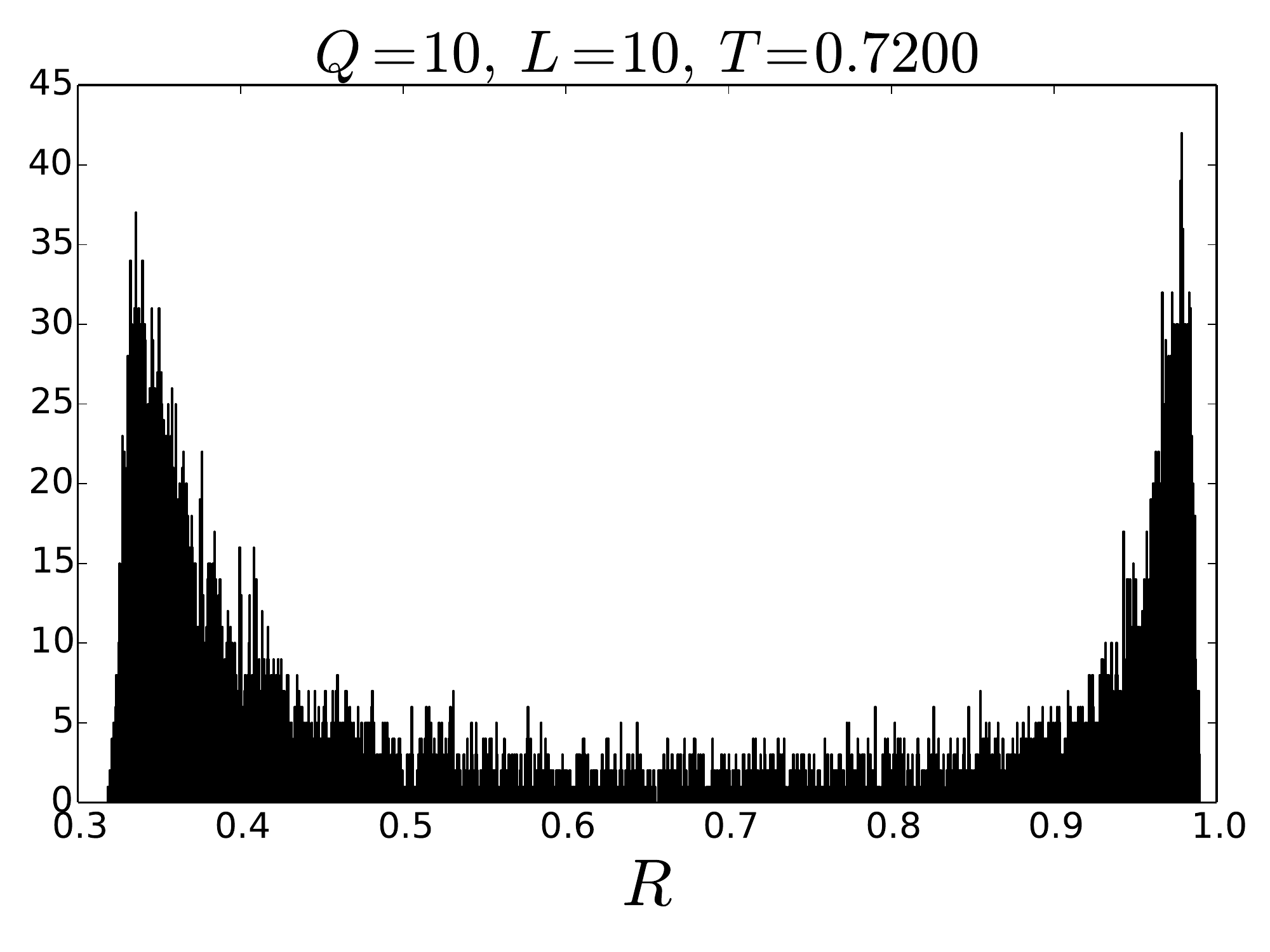}\vskip0.25cm
\includegraphics[width=0.325\textwidth]{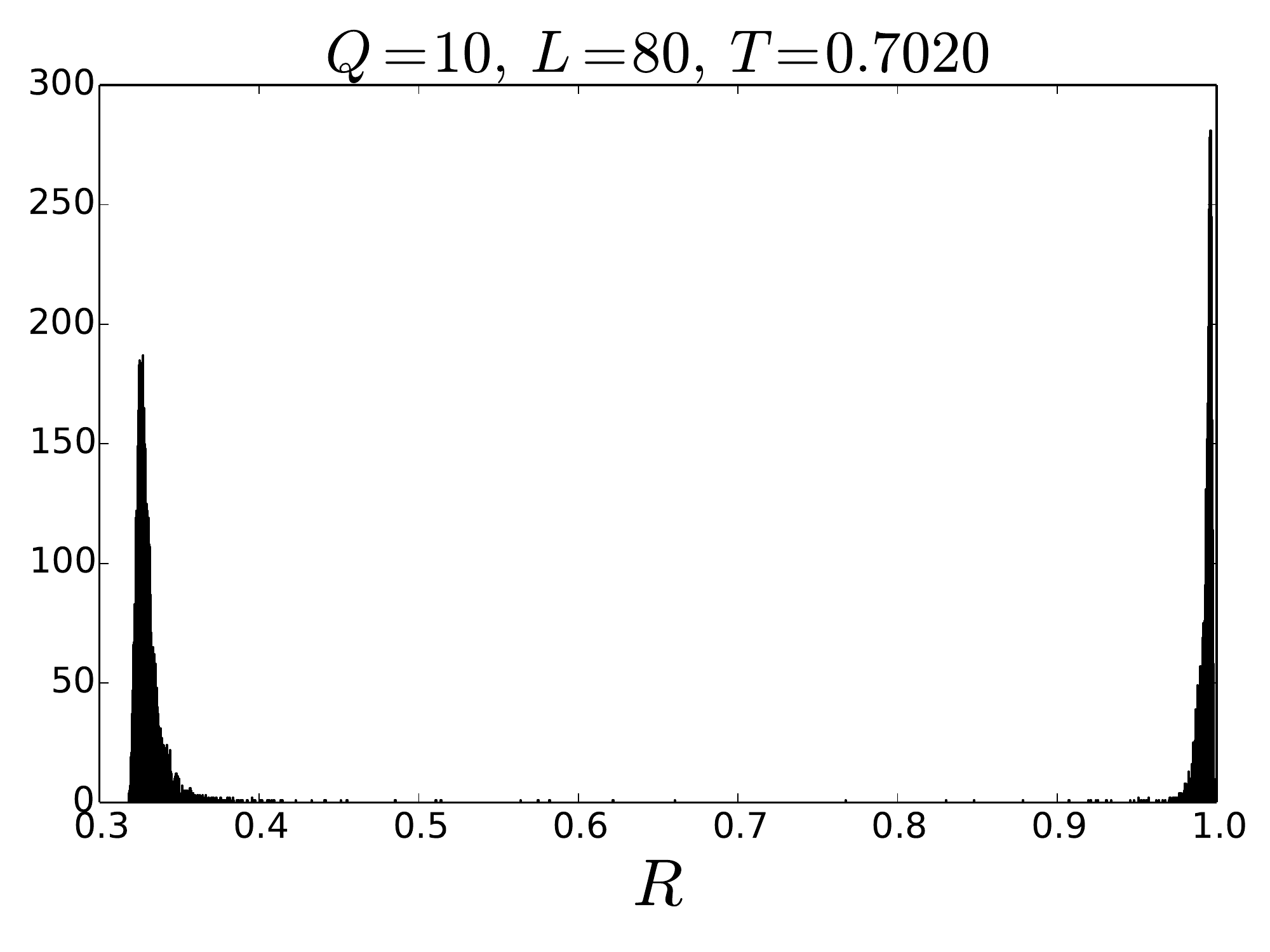}
}
\end{center}\vskip-0.7cm
\caption{Histograms of $R$ for $L = 10$ (top)
and 80 (bottom) for 2D $10$-states Potts model. The testing sets for these results 
are the same as those used in fig.~\ref{NN_2}. The outcomes shown here are obtained using 
different convolutional kernel than that considered in fig.~\ref{NN_2}. The early stop criterion
is removed in the calculations as well. The histograms 
are produced by the ``hist'' function of pylab \cite{pyl}.}
\label{conclusion_fig1}
\end{figure}

\section{Discussions and Conclusions}

\begin{figure}
\vskip0.5cm
\begin{center}
\vbox{
\includegraphics[width=0.325\textwidth]{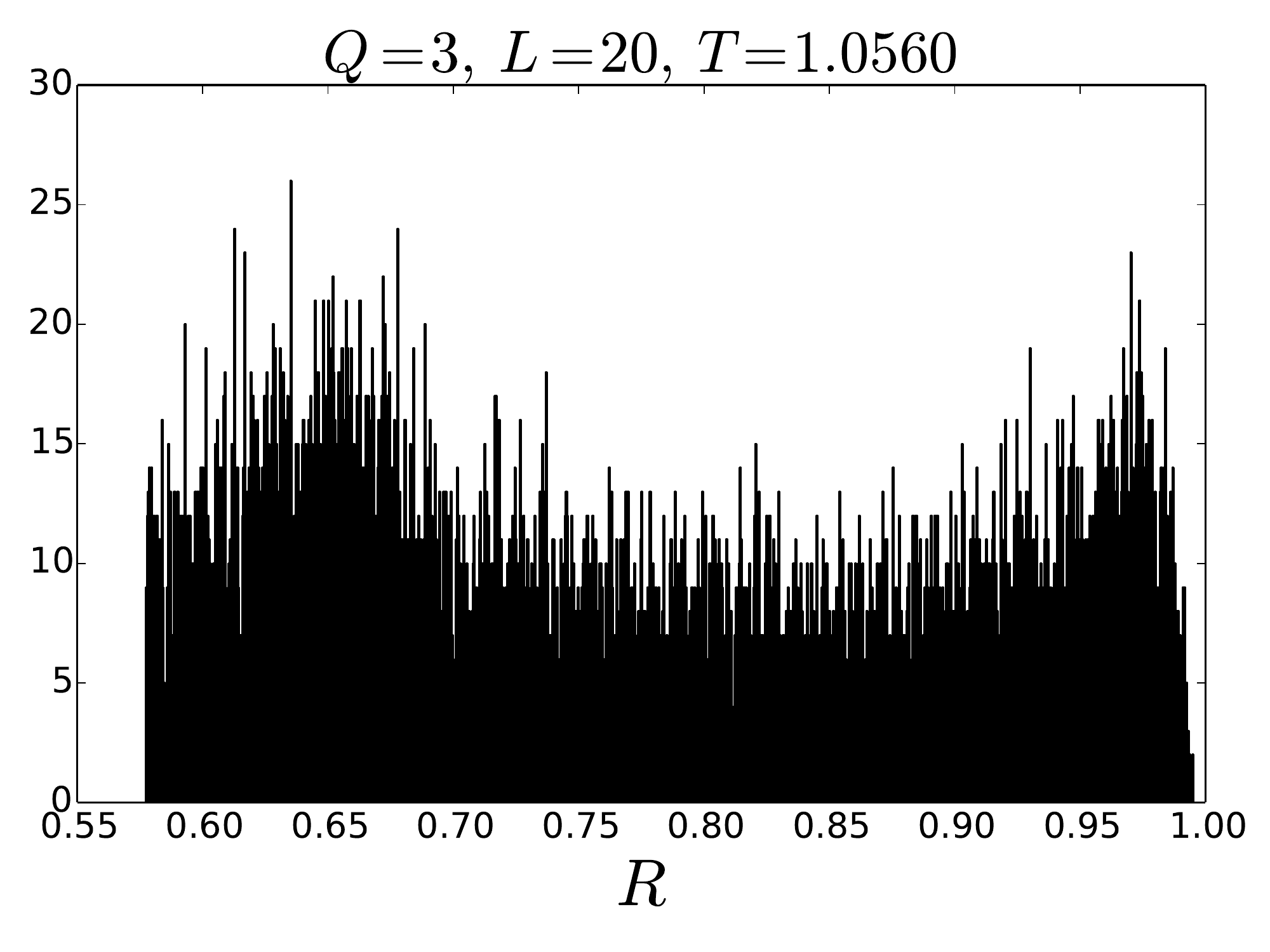}\vskip0.25cm
\includegraphics[width=0.325\textwidth]{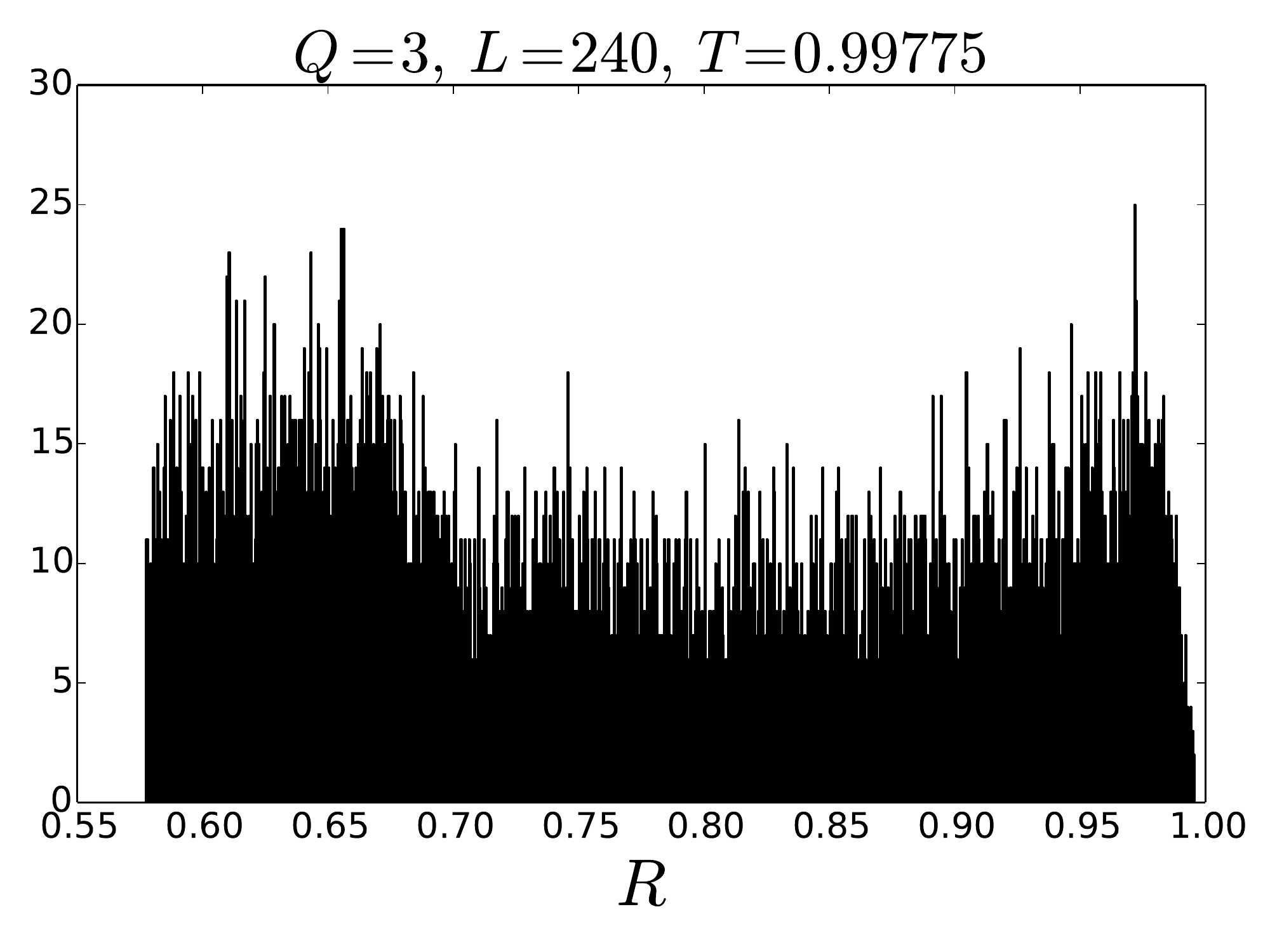}
}
\end{center}\vskip-0.7cm
\caption{Histograms of $R$ for $L = 20$ (top)
and 240 (bottom) for 2D $3$-states Potts model. The testing sets for these results 
are the same as those used in figs.~\ref{NN_3} and \ref{NN_4}. 
The outcomes shown here are obtained using 
different convolutional kernel than that considered in fig.~\ref{NN_3}. The histograms 
are produced by the ``hist'' function of pylab \cite{pyl}.}
\label{conclusion_fig2}
\end{figure}

%{\bf Discussions and Conclusions} --- 
In this study, we investigate the 
phase transitions of various 2D $Q$-states
Potts models using both the MC simulations and the techniques of NN. Special 
attention has been focused on whether the NN 
methods can be applied efficiently to determine the nature of the considered
phase transitions. Remarkably, while one expects MC simulations 
is capable of carrying out such tasks, it is surprising that the simple NN 
constructed here provides us with clear signals to decide whether these
phase transitions are first order or second order. Many pioneering works have 
demonstrated that NN is a powerful tool to distinguish phases of matters,
and the results shown here certainly strengthen the applicability of NN in 
investigating condensed matter systems. One natural next step will 
be to build a NN capable of accurately calculating physical quantities, such as the critical 
exponents, of second order phase transitions.     
     
In computer science like the artificial intelligence, the ideas behind 
designing algorithms are sophisticated so that the constructed algorithms
can apply for computations in many different areas. Such a concept is
well-suited in majority of scientific fields including physics. Notice different 
physical systems have their own special characters. Hence constructing
appropriate NN based on certain properties of the considered physical
systems is also an efficient way to conduct the investigation. 
In particular, several professional ML packages and libraries like Keras and TensorFlow are 
publicly available \cite{kera,scik,tens,thea}. These ML packages and libraries 
can be easily adopted to
study specific systems. Hence for systems with less complexity, 
the strategy considered here, namely designing the NN according to
the characters of the investigated models is an effective approach as 
well. We would like to emphasize the fact that unlike other supervised NN
used for studying the phase transitions of condensed matter systems,
in our investigation no configurations of the disordered phase are in
the pre-training set. This strategy greatly reduces the cost of the training
process. Indeed, we have considered the situation of including the disordered 
configurations and using only two numbers as the outputs. The training for
such trial calculations is much more demanding. In particular the related 
results are less accurate than those presented in the previous sections.
Finally, we would like to point out as well that to determine
$T_c$ using the ideas proposed in this study does not require any a priori
information of $T_c$. 

Notice a typical NN has several 
tunable parameters. For example, for the constructed NN here, 
the batch-size, the size of convolutional kernel, as well as the number of 
training objects are all adjustable variables. For the obtained results to
be reliable, one has to make sure that the conclusions stay the same when the 
tunable parameters are reasonably varied. For instance, we have tested the 
results determined by using filter kernels with 
different sizes and matrix elements. In addition, We have also considered
the situations of modifying the average pooling layer.
Figure \ref{conclusion_fig1} shows the histograms of $R$ for $L=10$ and 80 for 
10-states Potts model. 
In particular, in that calculations we use 3 by 3 convolutional kernel with 
each matrix element being randomly picked in $\left(-1/3,1/3 \right)$. The 
early stop criterion is removed as well. The two peaks phenomena still remain 
quite visible in 
fig.~\ref{conclusion_fig1}.
Similarly, for 3-states Potts model the histograms of $R$ for $L = 20$ and 
$240$, obtained using 2 by 2 convolutional kernel, are shown in 
fig.~\ref{conclusion_fig2}. 
The results presented in fig.~\ref{conclusion_fig2} lead to the same conclusion
as that determined earlier, namely the phase transition of 3-states Potts 
model is second order.  

In summary, based on the new analysis associated with the modified NN,
we find that while in the case of $L=240$ one may not be able to arrive at 
satisfactory outcomes \cite{L240}, the resulting histograms of $R$ greatly favor the
conclusions as those presented in previous sections regarding the nature of
these considered phase transitions, provided that the tunable parameters 
are reasonably varied. In other words, the built NN is quite robust.
Finally, we would like to emphasize the fact that 
although the conclusions regarding the nature of the considered phase
transitions are unchanged when the used NN is modified, the corresponding 
$T_c(Q,L)$ at finite $L$ may slightly vary. This effect can be
treated as one part of the errors for $T_c(Q,L)$ related to the NN methods.

\section*{Acknowledgement}
Partial support from Ministry of Science and Technology of Taiwan is 
acknowledged. We thank D.~Banerjee for reading the manuscript carefully
and giving us very useful comments.


\begin{thebibliography}{1}
\bibitem{histograms}
In this study, we do not carry out a detailed error analysis since very fine histograms
of the relevant observables are used for the investigation.
\bibitem{pyl}
All the histograms shown in this study are produced by the ``hist'' function of
pylab. For better visualization, each of the histograms presented here may be 
obtained using different number of bins.
\bibitem{L240}
Occasionally one would arrive at unsatisfactory results for $L = 240$ with the NN shown
in fig.~\ref{NN_1}.
\end{thebibliography}
\end{document}